\documentclass[a4paper,12pt,twoside]{article} 
\usepackage[english]{babel} \usepackage{graphicx}

\def\lsim{\raisebox{-.4ex}{\rlap{$\sim$}} \raisebox{.4ex}{$<$}}
\def\gsim{\raisebox{-.4ex}{\rlap{$\sim$}} \raisebox{.4ex}{$>$}}
\newcommand {\pom}  {I\hspace{-0.2em}P}
\newcommand {\xpom} {\mbox{$x_{_{\pom}}$}}

\usepackage{rotating}
\usepackage{epsfig}
\usepackage[centertags]{amsmath}
\usepackage{amssymb,amscd}
\usepackage{array}
\usepackage{multirow}
\usepackage{supertabular}
\usepackage{colordvi}
\usepackage{dcolumn}
\usepackage{cite}
\usepackage{xspace}
\def\zero{{\scriptscriptstyle 0}}

\def\Z0{\ensuremath{Z^\zero}}

\def\SU2U1{{\rm SU}(2)\times{\rm U}(1)}

\def\exp{{\rm exp}}

\mathchardef\qsm=63
\mathchardef\pls=43
\mathchardef\mns=512
\mathchardef\plm=518
\mathchardef\eql=61
\mathchardef\smallleft=300
\mathchardef\smallright=301
\mathchardef\perslsh=47
\mathchardef\les=316
\mathchardef\gre=318
\mathchardef\leq=532
\mathchardef\grq=533
\chardef\usc=95
\chardef\til=126

\def\sqr#1#2#3{{\vcenter{\hrule height.#3ex\hbox{\vrule width.#2ex height#1ex
    \kern#1ex\vrule width.#3ex}\hrule height.#2ex}}}

\def\angleto{\vrule width.035em height2.1ex depth-.56ex\unskip\kern-.6ex\to}
\def\perchc#1{{\raise.4ex\hbox{$\mkern4mu#1{\it\perslsh}_
             {\mkern-5mu\scriptscriptstyle{{\rm o}\!{\rm o}}}^
             {\mkern-12.8mu\scriptscriptstyle{\rm o}}$}}}

\catcode`\@=11 
\def\parenbar{\mathpalette\p@renb@r}
\def\p@renb@r#1#2{\vbox{%
  \ifx#1\scriptscriptstyle \dimen@.7em\dimen@ii.2em\else
  \ifx#1\scriptstyle \dimen@.8em\dimen@ii.25em\else
  \dimen@1em\dimen@ii.4em\fi\fi \offinterlineskip
  \ialign{\hfill##\hfill\cr
    \vbox{\hrule width\dimen@ii}\cr
    \noalign{\vskip-.3ex}%
    \hbox to\dimen@{$\mathchar300\hfil\mathchar301$}\cr
    \noalign{\vskip-.3ex}%
    $#1#2$\cr}}}
\catcode`\@=12 

\newbox\struttbox
\setbox\struttbox=\hbox{\vrule height1.65ex depth.485ex width0pt}
\def\strutt{\relax\ifmmode\copy\struttbox\else\unhcopy\struttbox\fi}
\def\stru#1#2{\relax\ifmmode\hbox{\vrule height#1 depth#2 width0pt}
\else\vrule height#1 depth#2 width0pt\fi}

\def\ronum#1{\uppercase\expandafter{\romannumeral#1}}
\def\ronuml#1{\expandafter{\romannumeral#1}}

\DeclareMathAlphabet{\mathbf}{OT1}{cmr}{bx}{sl}

\renewcommand{\arraystretch}{1.1}
 {\end{list}}
 {\end{list}}
 {\end{list}}

\catcode`\@=11 
\newlength{\@fninsert}
\newlength{\@fnwidth}
\renewcommand{\@makefntext}[1]%
  {\noindent\makebox[\@fninsert][r]{\@makefnmark}\hfil%
  \parbox[t]{\@fnwidth}{#1}}
\catcode`\@=12 
\newlength{\localtextwidth}
\catcode`\@=11 
\newsavebox{\tmpbox}
\newlength{\@captionmargin}
\newlength{\@captionwidth}
\newlength{\@captionitemtextsep}
\renewcommand{\@makecaption}[2]%
  {%
   \vspace{10.pt}
   \setlength{\@captionwidth}{\localtextwidth}
   \addtolength{\@captionwidth}{-\@captionmargin}
   \sbox{\tmpbox}{{\bf #1:}{\it #2}}%
   \ifthenelse{\lengthtest{\wd\tmpbox > \@captionwidth}}%
   {\centerline{\parbox[t]{\@captionwidth}%
   {\tolerance=2000\normalsize%
    {\bf #1:}\hspace{\@captionitemtextsep}{\it #2}}}}%
   {\centerline{{\bf #1:}\kern1.em{\it #2}}}}
\catcode`\@=12 
\catcode`\@=11 
\renewcommand\section{\@startsection{section}{1}{\z@}%
                                   {-3.5ex \@plus -1ex \@minus -.2ex}%
                                   {2.3ex \@plus.2ex}%
                                   {\normalfont\Large\bfseries}}
\renewcommand\subsection{\@startsection{subsection}{2}{\z@}%
                                   {-3.25ex\@plus -1ex \@minus -.2ex}%
                                   {1.5ex \@plus .2ex}%
                                   {\normalfont\large\bfseries}}
\renewcommand\subsubsection{\@startsection{subsubsection}{3}{\z@}%
                                   {-3.25ex\@plus -1ex \@minus -.2ex}%
                                   {1.5ex \@plus .2ex}%
                                   {\normalfont\large\bfseries}}
\renewcommand\paragraph{\@startsection{paragraph}{4}{\z@}%
                                   {3.25ex \@plus1ex \@minus.2ex}%
                                   {1.2ex \@plus .2ex}%
                                   {\normalfont\normalsize\bfseries}}
\catcode`\@=12 

\newsavebox{\sesbox}
\newlength{\seslen}

\begin{document}
\selectlanguage{english}
\thispagestyle{empty}

\bigskip
\bigskip

\title{
\vspace{1.5cm}
\bf\LARGE Measurement of the $Q^2$ and energy dependence 
of diffractive interactions at HERA
}                  
                    
\author{ZEUS Collaboration}
\date{}
\maketitle

\vspace{-9cm}                                     
\begin{flushright}
\tt DESY-02-029\\ 
 March 2002 
\end{flushright}

\bigskip
\bigskip
\bigskip
\bigskip
\vfill
\centerline{\bf Abstract}
\vskip4.mm
\centerline{
  \begin{minipage}{15.cm}
    \noindent
Diffractive dissociation of virtual photons, $\gamma^* p \rightarrow X
p$, has been studied in $ep$ interactions with the ZEUS detector at HERA. 
The data cover 
photon virtualities $0.17 < Q^2< 0.70$~GeV$^2$ and $3 < Q^2<
80$~GeV$^2$
with $3<M_X<38$ GeV, where $M_X$ is the mass of the hadronic final state.
Diffractive events were selected by two methods:
the first required the detection of the scattered proton in the ZEUS
leading proton spectrometer (LPS); the second was based on the
distribution of $M_X$. The integrated luminosities of the low- and
high-$Q^2$ samples used in the LPS-based analysis are 
$\simeq 0.9$~pb$^{-1}$ and $\simeq 3.3$~pb$^{-1}$, respectively. The
sample used for the $M_X$-based analysis corresponds to an 
integrated luminosity of $\simeq 6.2$~pb$^{-1}$.   
The dependence of the diffractive cross section
on $W$, the virtual photon-proton centre-of-mass energy, and on $Q^2$
is studied.
In the low-$Q^2$ range, 
the energy dependence is compatible with Regge theory
and is used to determine the intercept of the Pomeron trajectory.
The $W$ dependence of the diffractive cross section exhibits no
significant change from the low-$Q^2$ to the high-$Q^2$ region.
In the low-$Q^2$ range, little $Q^2$ dependence is
found, a significantly different behaviour from the rapidly falling cross
section measured for $Q^2 > 3$ GeV$^2$.
The ratio of the diffractive to
the virtual photon-proton total cross section is studied as a function of
$W$ and $Q^2$. Comparisons are made with a model based on perturbative
QCD.
  \end{minipage}
  }
\vfill

\newpage

%
%
%
%
\topmargin-1.cm                                                                                    
\evensidemargin-0.3cm                                                                              
\oddsidemargin-0.3cm                                                                               
\textwidth 16.cm                                                                                   
\textheight 680pt                                                                                  
\parindent0.cm                                                                                     
\parskip0.3cm plus0.05cm minus0.05cm                                                               
\def\3{\ss}                                                                                        
\newcommand{\address}{ }                                                                           
\pagenumbering{Roman}                                                                              
                                                   %
\begin{center}                                                                                     
{                      \Large  The ZEUS Collaboration              }                               
\end{center}                                                                                       
  S.~Chekanov,                                                                                     
  D.~Krakauer,                                                                                     
  S.~Magill,                                                                                       
  B.~Musgrave,                                                                                     
  A.~Pellegrino,                                                                                   
  J.~Repond,                                                                                       
  R.~Yoshida\\                                                                                     
 {\it Argonne National Laboratory, Argonne, Illinois 60439-4815}~$^{n}$                            
\par \filbreak                                                                                     
  M.C.K.~Mattingly \\                                                                              
 {\it Andrews University, Berrien Springs, Michigan 49104-0380}                                    
\par \filbreak                                                                                     
  P.~Antonioli,                                                                                    
  G.~Anzivino$^{   1}$,                                                                            
  G.~Bari,                                                                                         
  M.~Basile,                                                                                       
  L.~Bellagamba,                                                                                   
  D.~Boscherini,                                                                                   
  A.~Bruni,                                                                                        
  G.~Bruni,                                                                                        
  G.~Cara~Romeo,                                                                                   
  M.~Chiarini,                                                                                     
  L.~Cifarelli,                                                                                    
  F.~Cindolo,                                                                                      
  A.~Contin,                                                                                       
  M.~Corradi,                                                                                      
  S.~De~Pasquale,                                                                                  
  P.~Giusti,                                                                                       
  G.~Iacobucci,                                                                                    
  G.~Levi,                                                                                         
  A.~Margotti,                                                                                     
  T.~Massam,                                                                                       
  R.~Nania,                                                                                        
  C. Nemoz$^{   2}$,                                                                               
  F.~Palmonari,                                                                                    
  A.~Pesci,                                                                                        
  G.~Sartorelli,                                                                                   
  Y.~Zamora Garcia$^{   3}$,                                                                       
  A.~Zichichi  \\                                                                                  
  {\it University and INFN Bologna, Bologna, Italy}~$^{e}$
% ...
\newpage                                                                                           
$^{\    1}$ now at Universit\`a di Perugia, Dipartimento di                                        
Fisica, Perugia, Italy\\                                                                            
$^{\    2}$ now at E.S.R.F., Grenoble, France \\                                                    
$^{\    3}$ now at Inter American Development Bank, Washington DC, USA \\                           
% ...
                                                           %
                                                           %
\newpage   
                                                           %
                                                           %
\begin{tabular}[h]{rp{14cm}}                                                                       
$^{a}$ &  supported by the Natural Sciences and Engineering Research                               
          Council of Canada (NSERC) \\                                                             
$^{b}$ &  supported by the German Federal Ministry for Education and                               
          Research (BMBF), under contract numbers HZ1GUA 2, HZ1GUB 0, HZ1PDA 5, HZ1VFA 5\\         
$^{c}$ &  supported by the MINERVA Gesellschaft f\"ur Forschung GmbH, the                          
          Israel Science Foundation, the U.S.-Israel Binational Science                            
          Foundation, the Israel Ministry of Science and the Benozyio Center                       
          for High Energy Physics\\                                                                
$^{d}$ &  supported by the German-Israeli Foundation, the Israel Science                           
          Foundation, and by the Israel Ministry of Science\\                                      
$^{e}$ &  supported by the Italian National Institute for Nuclear Physics (INFN) \\                
$^{f}$ &  supported by the Japanese Ministry of Education, Science and                             
          Culture (the Monbusho) and its grants for Scientific Research\\                          
$^{g}$ &  supported by the Korean Ministry of Education and Korea Science                          
          and Engineering Foundation\\                                                             
$^{h}$ &  supported by the Netherlands Foundation for Research on Matter (FOM)\\                   
$^{i}$ &  supported by the Polish State Committee for Scientific Research,                         
          grant no. 620/E-77/SPUB-M/DESY/P-03/DZ 247/2000-2002\\                                   
$^{j}$ &  partially supported by the German Federal Ministry for Education                         
          and Research (BMBF)\\                                                                    
$^{k}$ &  supported by the Fund for Fundamental Research of Russian Ministry                       
          for Science and Edu\-cation and by the German Federal Ministry for                       
          Education and Research (BMBF)\\                                                          
$^{l}$ &  supported by the Spanish Ministry of Education and Science                               
          through funds provided by CICYT\\                                                        
$^{m}$ &  supported by the Particle Physics and Astronomy Research Council, UK\\                   
$^{n}$ &  supported by the US Department of Energy\\                                               
$^{o}$ &  supported by the US National Science Foundation\\                                        
$^{p}$ &  supported by the Polish State Committee for Scientific Research,                         
          grant no. 112/E-356/SPUB-M/DESY/P-03/DZ 301/2000-2002, 2 P03B 13922\\                     
$^{q}$ &  supported by the Polish State Committee for Scientific Research,                         
          grant no. 115/E-343/SPUB-M/DESY/P-03/DZ 121/2001-2002, 2 P03B 07022\\                      
\end{tabular}                                                                                      
                                                           %
                                                           %

\newpage

\pagenumbering{arabic}

\section{Introduction}

The properties of high-energy hadron-hadron cross sections, 
notably the energy dependence of the total and elastic cross sections,
are described successfully by Regge phenomenology
in terms of the exchange of the Pomeron trajectory, 
$\alpha_{I\hspace{-0.1cm}P}(t)
=\alpha_{I\hspace{-0.1cm}P}(0)+\alpha_{I\hspace{-0.1cm}P}'t$,
where $t$ is the squared four-momentum carried by the
exchange~\cite{pdbc}.
The intercept and slope of the trajectory were found to be 
$\alpha_{I\hspace{-0.1cm}P}(0)= 1.08$ and
$\alpha_{I\hspace{-0.1cm}P}'=0.25$\,GeV$^{-2}$, respectively,
by Donnachie and Landshoff \cite{dl}
using the energy dependence of the hadron-hadron total and elastic
cross sections.
Such a Pomeron trajectory is referred to as ``the soft Pomeron''.
At high energies, hadron-hadron total cross sections, including the 
$\gamma p$ total cross section, can be expressed in terms of this
trajectory as
\begin{displaymath}
        \sigma \propto (W^2)^{\alpha_{I\hspace{-0.1cm}P}(0)-1},
\end{displaymath}
\noindent 
where $W$ is the virtual photon-proton centre-of-mass energy.

Measurements of the diffractive dissociation of photons have shown that, 
for quasi-real photons ($Q^2 \approx 0$, photoproduction,
where $Q^2$ is
the exchanged photon virtuality), the value of
$\alpha_{\pom}(0)$ is compatible with the expectations 
based on soft-Pomeron exchange~\cite{h1mx,zeusphp}.
The study of diffractive processes in $ep$
collisions at large virtualities has opened up the possibility of
investigating the Pomeron
in a regime where perturbative QCD (pQCD) is applicable~\cite{review_halina}. 
In this regime, the exchange of the Pomeron trajectory may be
described, at lowest order, as two-gluon exchange in the
$t$ channel so that the cross section is proportional to the square of
the gluon density in the proton. Since the gluon
distribution rises steeply at small Bjorken $x$ (or, equivalently,
for large values of $W$), 
a possible signature of the 
transition from the soft non-perturbative regime
to the hard perturbative regime is a change to a 
$W$-dependence of the cross section steeper than that from the
exchange of a soft-Pomeron 
trajectory.
The value of the Pomeron intercept,
$\alpha_{\pom}(0)$, measured in the deep inelastic scattering 
(DIS) regime ($Q^2 \gsim$ a few GeV$^2$) is larger 
than that of the soft Pomeron \cite{h1diff,zeusdiff94}, which suggests
that pQCD
effects have become important.

In analogy with the usual DIS formalism for the proton structure function,
$F_2$,
one can introduce a diffractive
structure function, $F_2^D$.
Studies of photon diffractive-dissociation  have shown
that, for $Q^2~\gsim~1$~GeV$^2$, $F_2^D$ has only a weak, logarithmic, 
dependence on $Q^2$~\cite{h1diff,zeusdiff94}. 
However, conservation of the electromagnetic current requires that both 
$F_2^D$ and $F_2$ must behave like $Q^2$ as $Q^2 \rightarrow 0$.

In this paper, the inclusive diffractive dissociation of virtual 
photons, $\gamma^*p \rightarrow Xp$, is investigated by studying the
reaction  $ep \rightarrow eXp$ at HERA 
both in the perturbative region ($Q^2 \gg 1$~GeV$^2$) and in
the transition region between the non-perturbative ($Q^2 \sim 0$)
and perturbative regions. The measurements are presented as a function of 
$W$ and  $Q^2$.
The Pomeron intercept is determined  
through the measurement of the energy dependence of the 
diffractive cross section in the
transition region, which
has not previously been explored in diffraction. The $W$ and
$Q^2$ behaviour of the diffractive cross 
section and of the virtual photon-proton total cross section, 
$\sigma_{\rm tot}^{\gamma^*p}$,  are compared 
by studying their ratio as a function of $W$ and $Q^2$.

Diffractive events were selected by two methods.
The first required the detection of the
scattered proton in the ZEUS leading proton spectrometer (LPS) and is 
referred to as the ``LPS method''.
Although statistically limited because of the small acceptance of the LPS,
this method
permits the selection of events with negligible background from the
double-dissociative reaction, $ep \rightarrow eXN$, 
where the proton also diffractively dissociates into a state $N$ of mass 
$M_N$ that escapes undetected in the beam pipe.
The LPS method also gives access to higher values of $M_X$,
the mass of the hadronic final-state system, $X$,
and allows the measurement of the
squared four-momentum transfer at the proton vertex, $t$.
The second method, 
henceforth referred to as the ``$M_X^2$ method''~\cite{zeusdiff94},
is based on the characteristics of the distribution of $M_X$.
The sample selected  with the $M_X^2$ method
contains a background contribution from the
double-dissociative events.

This paper presents results in the region $0.17<Q^2<0.70$\,GeV$^2$, 
obtained
with both methods, and in the region $3<Q^2<80$~GeV$^2$ using only the LPS. 
 The measurements cover the region $3<M_X<38$ GeV. Results in
the DIS region obtained using the $M_X^2$ method have been previously
reported~\cite{zeusdiff94}.

\section{Kinematic variables and cross sections}
\label{s:kinrec}

Inclusive diffractive dissociation of virtual photons in positron-proton
collisions, $ep \rightarrow eXp$,
can be described by the kinematic variables $Q^2$, $W$, 
$M_X$, and $t$.
The differential cross section for $\gamma^*p \rightarrow X p$
is related to the cross section for the reaction $ep \rightarrow eXp$
by~\cite{flux}
\begin{equation}
        \frac{d^{4}\sigma^{ep}_{\rm diff}(Q^2, W, M_X, t)}{d\ln Q^{2}\,d\ln W\, dM_{X}\,dt}
        =
        \Gamma (Q^2, W)
        \frac{d^2\sigma_{\rm diff}^{\gamma^*p}(Q^2, W, M_X, t)}{dM_{X}\,dt},
        \label{diffcross1}
\end{equation}
where
\begin{displaymath}
        \Gamma=\frac{\alpha}{\pi}  \left[1+(1-y)^{2}\right]
\end{displaymath}
\noindent 
is the virtual photon flux, $\alpha$ is the fine-structure constant,
$y \simeq (W^2+Q^2)/s$ is the
fraction of the positron energy transferred to the proton in its rest
frame, and $s$ is the square of the positron-proton centre-of-mass energy.

In analogy with the formalism of inclusive deep inelastic $ep$ scattering,
the diffractive cross section for the reaction $ep \rightarrow eXp$ can also 
be expressed in terms of diffractive structure functions~\cite{ingelman}:
\begin{equation}
        \frac{d^4\sigma_{\rm diff}^{\rm ep}}{d\beta \,dQ^2 \,d\xpom \,dt} =
        \frac{4 \pi \alpha^2}{\beta Q^4} \; \left\{ 1 - y +
        \frac{y^2}
        {2(1 + R^{D(4)}(\beta,Q^2,\xpom,t))}
        \right\} \;
        F_2^{D(4)}(\beta,Q^2,\xpom,t),
        \label{first}
\end{equation}
where the diffractive structure function
$F_2^{D(4)}$ and the ratio of the cross sections for
longitudinal and transverse photons, $R^{D(4)}$,  have been introduced.

The variables $\xpom$ and $\beta$ are related to $Q^2$, $W^2$, $M_X^2$ and 
$t$ by
\begin{eqnarray}
        \xpom & = & \frac{Q^2+M_X^2-t}{Q^2+W^2-M_p^2}, \nonumber        
\label{rec-xpom}\\
        \beta & = & \frac{Q^2}{Q^2+M_X^2-t}, \nonumber
        \label{rec-beta}
\end{eqnarray}
\noindent
where $M_p$ is the proton mass. The variables $\xpom$ and $\beta$
can be interpreted,
assuming the $t$-channel exchange of a Pomeron with partonic structure,
as the fraction of the proton momentum carried by the Pomeron and the 
fraction of the Pomeron momentum carried by the struck parton, respectively.

Equations~(\ref{diffcross1}) and~(\ref{first}) can be combined to give
\begin{equation}
        \frac{d^{2}\sigma_{\rm diff}^{\gamma^*p}}{dM_X \, dt} =
        \frac{W^2}{Q^2+W^2} \,
        \frac{2 M_X}{Q^2+M_X^2} \,
        \frac{4\pi^2 \alpha}{Q^2} \,
        x_{I\hspace{-0.1cm}P}F_2^{D(4)},
        \label{f2d4b}
\end{equation}
where $|t| \ll Q^2+M_X^2$ has been assumed and  $M_p$ and $R^{D(4)}$ have
been neglected~\cite{rsmall}.
An analogous expression holds for the three-fold differential diffractive
structure function, $F_2^{D(3)}$, obtained by integrating $F_2^{D(4)}$
over $t$. Equation~(\ref{f2d4b}) is the diffractive analogue of the
expression
$\sigma_{\rm tot}^{\gamma^*p}=(4 \pi^2 \alpha/ Q^2)F_2$ which holds for 
inclusive $\gamma^*p$ scattering at high $W$.

\section{Experimental set-up }

The measurements were performed at the HERA $ep$ collider at DESY between
1995 and 1997
using the ZEUS detector. At that time, HERA operated at a proton energy of 
820~GeV and a positron energy of 27.5~GeV.

A detailed description of the ZEUS detector can be found
elsewhere~\cite{bluebook}. A brief outline of the components 
most relevant for this analysis is given below.

Charged particles are tracked by the central tracking detector (CTD)~\cite{ctd},
which operates in a magnetic field of 1.43\,T provided 
by a thin superconducting coil.
The CTD consists of 72 cylindrical drift-chamber layers,
organised in 9 superlayers covering the polar-angle\footnote{
The ZEUS coordinate system is a right-handed Cartesian system,
with the $Z$ axis pointing in the proton-beam direction,
referred to as the ``forward direction'', 
and the $X$ axis pointing left towards the centre of HERA.
The coordinate origin is at the nominal interaction point. 
The pseudorapidity is defined as $\eta = -\ln (\tan \frac{\theta}{2})$,
where the polar angle, $\theta$, is measured 
with respect to the proton-beam direction.}
region \mbox{$15^\circ < \theta < 164^\circ$.}
The relative transverse-momentum resolution for full-length tracks 
is $\sigma(p_t)/p_t=0.0058p_t\oplus 0.0065 \oplus 0.0014/p_t$,
with $p_t$ in GeV.

The high-resolution uranium-scintillator calorimeter (CAL)~\cite{cal} 
consists of three parts:
the forward (FCAL), the barrel (BCAL) and the rear (RCAL) calorimeters.
Each part is subdivided transversely into towers
and longitudinally into one electromagnetic section (EMC)
and either one (in RCAL) or two (in FCAL and BCAL) 
hadronic sections (HAC).
The smallest subdivision of the calorimeter is called a cell.
The CAL relative energy resolutions, as measured under test-beam conditions,
are $\sigma(E)/E=0.18/\sqrt{E}$ for electrons
and $\sigma(E)/E=0.35/\sqrt{E}$ for hadrons ($E$ in GeV).

Low-$Q^2$ events  ($0.17 <Q^2 < 0.70$~GeV$^2$) were tagged by requiring 
the identification of the scattered positron 
in the beam pipe calorimeter (BPC) \cite{zeusbpc,zeusbpt}.
The BPC was a tungsten-scintillator sampling calorimeter,
located close to the beam pipe,
3\,m downstream of the interaction point
in the positron-beam direction.
The relative energy resolution from test-beam results
was $\sigma(E)/E=0.17/\sqrt{E}$ ($E$ in GeV).
Each scintillator layer consisted of 8\,mm-wide strips.
Using the logarithmically weighted shower position, 
the impact position of the scattered positron 
could be measured with an accuracy of about 1\,mm.
For  events  with $Q^2>3$~GeV$^2$, the impact point of the scattered 
positron was determined with the small-angle rear tracking detector 
(SRTD)~\cite{srtd} or the CAL. The SRTD
is attached to the front face of the RCAL and consists of two planes of 
scintillator strips, 1~cm wide and 0.5~cm thick, arranged in orthogonal 
orientations
and read out via optical fibres and photomultiplier tubes. It covers a
region of about $68 \times 68$~cm$^2$ in $X$ and $Y$, excluding a
$10 \times 20$~cm$^2$ hole at the centre for the beam pipe. 

The LPS~\cite{lpsrho} detected  positively
charged particles scattered at small angles and carrying a substantial
fraction, $x_L$, of the incoming proton momentum; these particles remain
in the beam pipe and their trajectory was measured by a system of
silicon microstrip detectors that can be inserted 
very close (typically a few mm)
to the proton beam. The detectors were grouped in six stations,
S1 to S6, placed along the beam line in the direction of the 
proton beam, between $23.8$~m and 90.0~m from the interaction point.
The track deflections induced by the magnets of the
proton beam-line allow a momentum analysis of the scattered proton.
For the present measurements, only stations S4, S5 and S6 were used.
The resolutions were better than 1\% on the longitudinal momentum and 5~MeV on
the transverse momentum. The effective transverse-momentum
resolution is, however, dominated
by the intrinsic transverse-momentum spread of the proton beam at the
interaction point, which is about 40 MeV in the horizontal plane and
about 100 MeV in the vertical plane.

\section{Reconstruction of the kinematic variables}
Different methods have been used for the reconstruction of the kinematic
variables $Q^2$ and $W$, depending on the $Q^2$ range of the measurement.
At low $Q^2$, $0.17 <Q^2 < 0.70$~GeV$^2$ (hereafter referred to as
the ``BPC region''), the energy, $E_e'$, and angle,
$\theta_e$, of the scattered positron measured in
the BPC were used (``electron method'') to determine the kinematic
variables from
\begin{eqnarray*}
                Q^2 & = & 2E_eE_e'(1+\cos\theta_e), \\
                W   & = & \sqrt{4 E_e E_p \left[
                        1 - {\displaystyle \frac{E_e'}{2E_e}}
                        (1-\cos \theta_e)
                \right]},
        \label{e:ELMETHOD}
\end{eqnarray*}
where $E_{p}$ and $E_{e}$ represent the proton and positron beam energies,
respectively.
For $Q^2>3$~GeV$^2$ (the ``DIS region''), $Q^2$ and $W$
were reconstructed with the double angle method~\cite{DA} using the
energy depositions in the CAL.

For the reconstruction of the mass of the diffractive system $X$,
the energy deposits in the CAL
and the track momenta measured in the CTD were clustered to form 
energy-flow objects (EFOs)~\cite{zeusdiff94,gb}. The EFOs thus include 
the information from both neutral and charged particles in an optimal way.
The mass, $M_X$, was then
obtained from the EFOs via
\begin{equation}
        M_{X}=\sqrt{\left(\sum_{i}E_i\right)^2
-\left(\sum_{i}p_{Xi}\right)^2
            -\left(\sum_{i}p_{Yi}\right)^2
-\left(\sum_{i}p_{Zi}\right)^2}, \nonumber
        \label{rec-mx}
\end{equation}
where the subscript $i$ denotes an individual EFO; the EFOs associated with
 the scattered positron are excluded from the sums.

The momentum of those scattered protons detected in the LPS, $p^{\rm LPS}$, 
was measured, along with its component perpendicular (parallel) to the beam 
direction,
$p_T^{\rm LPS}$ ($p_Z^{\rm LPS}$).
From these quantities,
the fractional momentum of the scattered proton, $x_L$, and $t$ 
were determined via
\begin{eqnarray*}
        x_L & = & p_Z^{\rm LPS}/E_p, \\
        t  & = &  -\frac{(p_T^{\rm LPS})^2}{x_L}. \nonumber
        \label{eqt}
\end{eqnarray*}

Two quantities, $y$ and 
$\delta \equiv \sum_{i}(E-p_{Z})_{i}+E_{e'}(1-\cos\theta_e)$, the sum
of $E-P_Z$ over all final-state particles in the event,
were used in the event selection.
The former was reconstructed either using the electron method (and denoted by
$y_e$) or from the EFOs using the Jacquet-Blondel estimator~\cite{jb} as
\begin{equation}        
        y_{JB}=\frac{\sum_{i} (E-p_{Z})_{i}}{2E_{e}}, \nonumber
        \label{rec-yjb}
\end{equation}
where the sum is over all EFOs, excluding those assigned to 
the scattered positron. Energy and momentum conservation require $\delta $
to be twice the positron beam energy for a completely measured final
state and neglecting resolution effects.

\section{Monte Carlo simulation }
Monte Carlo (MC) generators were used to determine the acceptance of 
the apparatus.
The reaction $ep \rightarrow eXp$ was simulated in the BPC region with the
EPSOFT2.0~\cite{zeusdiff94,mk,masahide} MC generator
interfaced to HERACLES4.6~\cite{heracles}, 
which simulates initial- and final-state QED radiation. For the 
description of the 
diffractive dissociation of virtual photons, $\gamma^{*}p \rightarrow Xp$,
EPSOFT uses  the triple-Regge formalism~\cite{pdbc}, in which
the inclusive diffractive cross section can be expressed
in terms of three trajectories.
If all the trajectories are Pomerons 
($I\hspace{-0.1cm}PI\hspace{-0.1cm}PI\hspace{-0.1cm}P$),
the cross-section $d\sigma/dM_{X}^{2}$ 
is approximately proportional to $1/M_{X}^{2}$.
If one of the trajectories is a Reggeon 
($I\hspace{-0.1cm}PI\hspace{-0.1cm}PI\hspace{-0.1cm}R$), 
the cross-section $d\sigma/dM_{X}^{2}$ 
falls as $\sim 1/M_{X}^{3}$. EPSOFT also simulates exclusive
vector-meson production, $ep \rightarrow eVp$, where $V=\rho^0$, $\omega$ or
$\phi$, and non-diffractive $ep$ interactions, $ep \rightarrow eY$. 
Production of $J/\psi$ mesons has negligible effects on the acceptance and
was not considered.
EPSOFT was also used to simulate the double-dissociative reaction, 
$ep \rightarrow eXN$, where the proton diffractively dissociates into the state $N$.

The second generator, used for the DIS region, was RAPGAP2.06~\cite{rapgap}, 
where, for the diffractive structure function, a factorisable expression
was assumed based on the model of Ingelman and Schlein~\cite{ingsch}. 
In particular, a superposition of non-interfering Pomeron and sub-leading
trajectories was used 
(``fit B", as determined by the H1 Collaboration~\cite{h1diff}) together with the ``fit 
3" 
Pomeron parton density functions~\cite{h1diff}.
Again, initial- and final-state QED radiation were simulated
using HERACLES. 

All generated events were passed through the standard ZEUS detector
simulation, based on the GEANT3.13 program~\cite{geant},
and the trigger-simulation package.

\section{Identification of the scattered positron}
\label{scatt_e}

For the BPC sample, the events were selected in the trigger by requiring the 
presence of a scattered positron in the BPC. A
positron with energy greater than 7~GeV was required 
in the offline analysis~\cite{zeusbpc,zeusbpt}.
The following cuts were applied to reduce the contamination from photoproduction
events, radiative events, and beam-related background:
\begin{itemize}
        \item $y_{JB}>0.05$;
        \item $35<\delta<65$~GeV;
        \item $|Z_{\rm VTX}|<50$~cm, where $Z_{\rm VTX}$ is the $Z$ coordinate of the
            reconstructed vertex.
\end{itemize}

For the DIS sample, the events used for the analysis were selected in
the trigger by requiring the presence of a scattered positron in the
CAL.  Offline, a positron in the RCAL
with energy greater than 10~GeV was required. A positron finder based on
a neural-network was used~\cite{sinkus}.
The following cuts were applied to reduce the contamination from photoproduction
events, radiative events, and beam-related background:
\begin{itemize}
        \item $y_{JB}> 0.03$;
        \item $y_{e}<0.95$;
        \item $35<\delta<65$~GeV;
        \item $-50 <Z_{\rm VTX}< 100$~cm. 
        \end{itemize}

\section{The LPS method}

Diffractive events are characterised by a final state proton scattered at
very small angle and with energy nearly equal to that of the incoming
proton. In the LPS method, diffractive events are then defined as those
having
a proton detected in the LPS with $x_L \approx 1$. Figure~\ref{fig:lnmx}(a) 
shows the measured $x_L$ spectrum, uncorrected for acceptance. The diffractive 
peak is clearly visible for values of $x_L$ close to unity. 
For the present analysis, $x_L>0.97$ was required. Previous
studies~\cite{lps_f2d3} indicate that the double-dissociative contribution
to such events is negligible.

Two data samples, collected in 1995, were analysed with
the LPS method. The BPC sample, corresponding to a luminosity of 
$0.90\pm 0.01$~pb$^{-1}$, covers the range $0.17< Q^2< 0.70$~GeV$^2$ and 
$90<W<250$~GeV. The DIS sample covers the region 
$3 < Q^2< 80$~GeV$^2$ and $80<W<250$~GeV, and corresponds to a luminosity of
$3.30\pm0.03$ pb$^{-1}$.
The analysis was limited to the range $3<M_X<38$~GeV for the BPC sample and
$3<M_X<33$~GeV for the DIS sample.

The candidate proton was tracked along the proton beam line and was
rejected if, at any point, the reconstructed minimum distance of approach
to the beam pipe, $\Delta_{\rm pipe}$, was less than $400~\mu$m or if the 
distance to the edge of the sensitive region of any LPS station,
$\Delta_{\rm plane}$, was
smaller than $200~\mu$m. These cuts reduce the sensitivity of the
acceptance to the uncertainty in the position of the beam-pipe apertures
and of the detector edges.
In addition, $t$ was required to be in the region
$0.073 <|t|<0.4$~GeV$^2$, where the LPS acceptance is well 
understood~\cite{lpsrho}.
Beam-halo background results from a scattered proton, with energy close to 
that of the beam, originating from an interaction of a beam proton with the 
residual gas or with the beam collimators. In this case, 
the proton measured in the LPS is uncorrelated with the activity in the 
central 
ZEUS detector. This background was suppressed by the requirement that the 
sum of the energy and the longitudinal component of the total 
momentum measured 
in the CAL, the BPC and the LPS be less than the kinematic limit of $2E_p$: 
$(E+p_Z)_{\rm CAL}+ (E+p_Z)_{\rm BPC} +2 p_Z^{\rm LPS}
<1665$~GeV. This cut takes into 
account the resolution of the measurement of $p_Z^{\rm LPS}$.
A residual beam-halo background of approximately 8\%, 
remaining after the cut, was subtracted statistically.

In the BPC analysis, the measured number of diffractive events was 
corrected bin-by-bin.
From this acceptance-corrected number of events, the cross section
for the diffractive dissociation of virtual photons,
$\gamma^*p \rightarrow X p$, was determined taking into account
the integrated luminosity, bin widths, and bin-centring corrections.

In the DIS analysis,
the cross section for the diffractive dissociation of virtual photons
at a given point within a bin
was obtained from the ratio
of the measured number of diffractive events
to the number of events in that bin predicted from
the MC simulation,
multiplied by the $\gamma^*p \rightarrow X p$ cross section calculated
at that point by the Monte Carlo generator. 
Both the acceptance and the bin-centring corrections were thus
taken from the MC simulation.

In both the BPC and DIS analyses,
the diffractive cross-section
$d\sigma_{\rm diff}^{\gamma^*p}/dM_{X}$ was directly measured
only in the region $0.073 <|t|<0.4$~GeV$^2$ and extrapolated
to the full $t$ range using the $t$ dependence assumed in the Monte Carlo
generator. In the region covered by the present measurements,
this is roughly equivalent to carrying out an integration over
$t$ assuming an exponential dependence on $t$,
$e^{-b|t|}$, with $b \sim 7.5$~GeV$^{-2}$.

\section{The \protect\boldmath{$M_X^2$} method}

Diffractive photon dissociation, $\gamma^* p \rightarrow Xp$, is
characterised by the exchange of a colourless object, the Pomeron, 
between the virtual photon and the proton.
This suppresses QCD radiation, and hence the production of hadrons, in the 
rapidity region between the hadronic system $X$ and the scattered proton, 
yielding a forward rapidity gap, a characteristic feature of diffractive
interactions. This feature is reflected in the dependence of the
cross section on $M_X$, 
$d\sigma^{\gamma^*p}_{\rm diff}/dM_X \propto 1/M_X^{2\alpha_{\pom}(0)-1}$,
i.e.
approximately flat as a function of $\ln{M_X^2}$.
In contrast, for non-diffractive events, large rapidity gaps are 
exponentially suppressed
by QCD radiation, which populates the region between the struck 
quark and the coloured proton remnant. In this case, under the assumption of 
uniform, random and uncorrelated particle emission in rapidity,
the $\ln M_{X}^{2}$ distribution falls exponentially towards low $M_X$
values. The different properties of the $\ln M_{X}^{2}$ distribution for
diffractive and non-diffractive events are
exploited in the $M_X^2$ method~\cite{zeusdiff94}.

The $M_X^2$ method was used to analyse BPC data taken in 1996-97, 
corresponding to an integrated luminosity of $6.2\pm 0.1$ pb$^{-1}$.
The kinematic range used was $0.22 <Q^2 < 0.70$~GeV$^2$, $90
<W<220$~GeV and $3.0<M_X<12.2$ GeV.

\subsection{Selection of the diffractive signal}
\label{selectionmx}

\begin{figure}

\begin{center}\includegraphics[clip,scale=0.80]{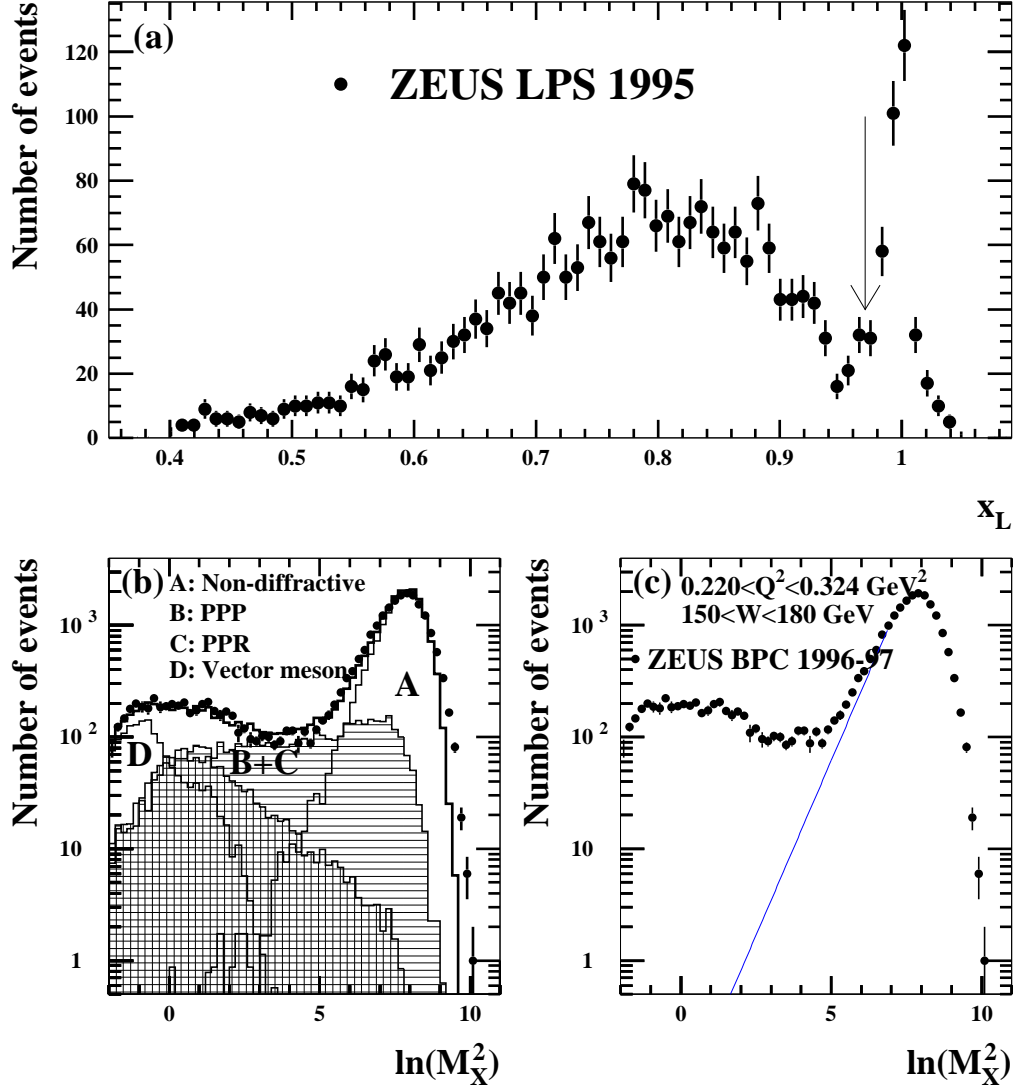}\end{center}
        \caption{
        (a) The $x_L$ spectrum as measured in BPC events with the LPS;
            (b) and (c) the $\ln M_{X}^{2}$ distribution ($M_X$ in {\rm GeV}) 
        of the  
        BPC data in the region $0.220 <Q^2 <0.324$~{\rm GeV}$^2$ and
       $150<W<180$~{\rm GeV}.
        In (a), the position of the arrow indicates the value 
$x_L=0.97$ used in the selection.
        In (b), the data are compared to the mixture of
                four kinds of EPSOFT MC events described in the text:
region A corresponds to non-diffractive events, B+C to the sum of the 
$I\hspace{-0.1cm}PI\hspace{-0.1cm}PI\hspace{-0.1cm}P$ and 
$I\hspace{-0.1cm}PI\hspace{-0.1cm}PI\hspace{-0.1cm}R$ 
contributions and D to the vector-meson contribution.
        In (c), the straight line shows the exponential slope,
        resulting from the fit described in the text, for non-diffractive
        events.
}
\label{fig:lnmx} \end{figure}

Figure~\ref{fig:lnmx}(b) shows a representative distribution of 
$\ln M_{X}^{2}$ 
for data\footnote{The data shown in Figs.~\ref{fig:lnmx}(b,c) result from
the cuts discussed in Section~\ref{pdissmx}.}
in the bin $0.220<Q^2<0.324$\,GeV$^2$ and $150<W<180$\,GeV,
compared to the distribution of the simulated events generated 
with EPSOFT.
Also shown are the four individual contributions generated with EPSOFT
for non-diffractive events,
for the 
$I\hspace{-0.1cm}PI\hspace{-0.1cm}PI\hspace{-0.1cm}P$ and 
$I\hspace{-0.1cm}PI\hspace{-0.1cm}PI\hspace{-0.1cm}R$ 
contributions (shown combined in the figure), which
lead to the diffractive dissociation of the photon, 
and for vector-meson production.
Diffractive events dominate the region of low $\ln M_{X}^{2}$,
while non-diffractive events exhibit a large peak at high $\ln M_{X}^{2}$ and
a steep exponential fall-off towards lower $\ln M_{X}^{2}$ values.
The relative weights of the four subprocesses are obtained from fits
to the $\ln M_{X}^{2}$ distribution of the data.
The resulting sum of the MC events (open histogram)
from the various subprocesses
provides a reasonably good description of the data
in the region of interest, $\ln M_{X}^{2}<8.5$.

In the region $\ln M_{X}^{2}\sim 4$, the diffractive contribution to the 
events in Fig.~\ref{fig:lnmx}(b) depends only weakly on 
$\ln M_{X}^{2}$. The expression
\begin{equation}
        \frac{dN}{d\ln M_{X}^{2}}=D+C~\exp (B\ln M_{X}^{2}) 
        \label{lnmx-fitfunc}
\end{equation}
was therefore fitted to the data and the parameters $D$, $C$ and $B$ were 
determined for each ($Q^2$,$W$) region. The exponential term (with 
$B=1.44 \pm 0.02$), ascribed to 
non-diffractive events, was subtracted statistically
to obtain the diffractive contribution. The parameter $D$ was thus
not used directly to determine the diffractive cross section. 
The exponential term resulting from the fit to the data of 
Fig.~\ref{fig:lnmx}(b) is shown in Fig.~\ref{fig:lnmx}(c).  

The cross-section measurement was restricted to the range $2.2<\ln M_X^2<5.0$,
corresponding to $3.0<M_X<12.2$\,GeV.
The lower limit on $M_X$ suppresses the contribution from
diffractive vector-meson production, while
the upper bound was chosen such that the non-diffractive
contribution to the higher-$M_X$ bins was always less than 50\%.

\subsection{Proton-dissociative contribution}
\label{pdissmx}

The diffractive sample of $ep \rightarrow eXp$ events
selected with the $M_X^2$ method
as discussed in Section~\ref{selectionmx} 
contains a contribution from the double-dissociative reaction 
$ep \rightarrow eXN$.
The system $N$ escapes undetected through the forward beam pipe, unless
the proton dissociates into a state of sufficiently high
mass, in which case some of the particles from the system $N$ have transverse
momenta large enough that they are detected in the FCAL 
region around the forward beam pipe.
The contribution of the double-dissociative reaction $ep \rightarrow eXN$ was
simulated and studied with EPSOFT.

Energy deposits in the FCAL, arising from the proton-dissociative remnant, 
give rise to a measured value of $M_X$ considerably
higher than the true value. 
In such events, there is a gap in rapidity
between the FCAL deposits from the proton remnant at high $\eta$
and the system $X$ at lower $\eta$, and
the invariant mass of the low-$\eta$ system
is small with respect to the measured (apparent) $M_X$.
Exploiting these features, events were rejected if they fulfilled all of
the following three conditions:
\begin{itemize}
 \item the maximum $\eta$ ($\eta_{\rm MAX}$) of the EFOs was
       greater than 2.5;
 \item the maximum rapidity gap between adjacent EFOs was
       greater than 3.5; 
 \item the mass reconstructed from EFOs 
       with $\eta <2.5$ was less than $0.6M_X$.
\end{itemize}
These cuts rejected approximately 10\% of the data sample. The simulations
using the EPSOFT MC program indicate that about 45\% of these 
rejected events are from the 
double-dissociative reaction, $\gamma^*p \rightarrow XN$, 
and the events that survive the cuts consist of
photon-dissociative events as
well as events from the double-dissociative reaction 
with $M_N < 6$~GeV.

The measured number of events in each $(Q^2,W,M_X)$  bin
was corrected for acceptance to determine the number of produced
events by means of the Singular Value Decomposition (SVD) 
method~\cite{svd}, which allows the evaluation of error correlations between 
adjacent bins.
The number of events thus obtained was divided
by the luminosity and
the bin-widths to evaluate the average
$ep \rightarrow eXp$ three-fold differential cross section
for each ($Q^2$,~$W$,~$M_X$) bin.
From this, the cross section
was obtained using Eq.~(\ref{diffcross1}), integrated over $t$ and
evaluated at the logarithmic centre of the bin.
The residual double-dissociative contribution in the data
leads to an overestimation of the cross section for the
diffractive dissociation of virtual photons, $\gamma^* p \rightarrow X p$;
it was evaluated using the LPS data and subtracted as discussed in 
Section~\ref{pdiss}.

\section{Systematic uncertainties}
\label{systematics}

The main sources of systematic uncertainty can be classified 
into three groups: the positron measurement in the BPC or the CAL; the
measurement of 
the hadronic final state in the CAL; and the measurement of the scattered proton
in the LPS:

\begin{itemize}

\item measurement of the scattered positron:

\begin{itemize}

\item 
for the BPC samples, the effects of the uncertainty in 
the absolute BPC energy scale~\cite{zeusbpt} ($\pm 0.5$\%), the
positron-selection criteria and the alignment of the BPC result in an 
uncertainty in the cross section that is
typically $\pm$7\% and always smaller than $\pm$20\%;

\item for the DIS sample, the 10 GeV cut on the 
scattered-positron energy was changed by $\pm 2$~GeV~\cite{lps_f2d3}. The
parameters of
the 
neural-network positron finder 
were modified. To check the acceptance at low $Q^2$, which is determined by the positron position in the SRTD, the fiducial
region around the impact point of the positron was changed. 
The resulting systematic uncertainty is typically
$\pm$7\% and always smaller than $\pm$25\%.

\end{itemize}

\item for the determination of the uncertainties related to the hadronic
final state for the $M_X^2$ method, the effect of the uncertainty in the CAL 
energy scale ($\pm 2$\%) was studied and the parameters in the algorithm that forms
EFOs were varied. The non-diffractive slope ($B$ in
Eq.~(\ref{lnmx-fitfunc})) was varied between 1.42 and 1.46. It was
checked that a different choice for the functional form of the
diffractive contribution in Eq.~(\ref{lnmx-fitfunc}) does not significantly
affect the final number of diffractive events. 
The resulting 
uncertainty is typically $\pm$8\% and always smaller
than $\pm$12\%;

\item
the systematic uncertainties
in the measurement of the scattered proton in the LPS were
estimated as follows:
\begin{itemize}
\item
to estimate the sensitivity of the LPS acceptance to the uncertainties in the
positions of the beam-line apertures, the lower limits on the distance of
closest approach to any of the beam-line elements and
to the edge of the sensitive region of each detector were raised from 400
$\mu$m to 1000~$\mu$m and from 200$~\mu$m to 300~$\mu$m, respectively;
\item
the $x_L$ window was varied by $\pm 0.01$;
\item
the uncertainty in the subtraction of the beam-halo events
was estimated by removing the $E + p_Z$ cut.
\end{itemize}
The resulting systematic uncertainty arising from the LPS
 measurement is
typically $\pm$10\% and always smaller than $\pm 25\%$.

\end{itemize}

In addition, the $M_X$, $W$ and $t$ dependences in EPSOFT and RAPGAP were
varied within the limits allowed by the data, yielding changes in the
cross section negligible with respect to all other uncertainties.
The relative fraction of vector meson production in EPSOFT was varied by
up to $\pm 10\%$, again with negligible effects on the results.

All the above contributions were summed in quadrature to give the final
systematic uncertainties. The normalisation
uncertainty due to the luminosity determination
is $\pm$1\% for the 1995 data and $\pm$1.5\%
for the 1996-97 data and was not included in the sum.

The value of $R^D$ was assumed to be zero throughout the analysis. Given
the absence of experimental information on $R^D$, no
attempt was made to quantify the systematic uncertainty entailed by this
assumption.

\section{Cross-section measurements and comparison 
of the \protect\boldmath{{$M_X^2$}} and LPS methods}
\label{pdiss}

The values of $d^{2}\sigma_{\rm diff}^{\gamma^*p}/dM_{X} dt$ and
$d\sigma_{\rm diff}^{\gamma^*p}/dM_{X}$
extracted with the LPS method
are given in Tables~\ref{tbl:lps1} and \ref{tbl:lps2}, respectively.
The results obtained with the $M_X^2$ method are presented in 
Table~\ref{tab:xsection}. All results are corrected to the Born level.

As discussed in Section~\ref{pdissmx}, the sample from the $M_X^2$ method 
contains a double-dissociative contribution.
Since the sample selected with the LPS method has a negligible 
proton-dissociative background~\cite{lps_f2d3}, the contamination in the
BPC sample was
estimated by directly comparing the results from 
the two analysis methods.

To achieve this, the ratio, $R_{M_X}$,
of the average cross section measured with the $M_X^2$ method
and that measured with the LPS method was determined in a single
kinematic region corresponding to the bins given in
Table~\ref{tab:xsection}.
The value obtained, $R_{M_X}= 1.85 \pm 0.38$~(stat.), is attributed to a
substantial contribution from the double-dissociative reaction to
the cross section measured with the $M_X^2$ method.
In terms of the ratio of the number of double-dissociative
events to the total number of events in the sample,
$R_{\rm diss}=(1-1/R_{M_X})$, the estimated value of $R_{M_X}$ corresponds to
$R_{\rm diss}=(46\pm 11)\%$.
This is consistent with
a previous estimate at higher $Q^2$ of
$(31 \pm 15$)\%~\cite{zeusdiff94}.

The results obtained with the $M_X^2$ method presented in this paper
were corrected for the residual double-dissociative background using the
measured value of $R_{M_X}$. The correction was assumed to
be independent of $W$ and $Q^2$, in agreement with the hypothesis of vertex
factorisation~\cite{vf}. The values of $d\sigma_{\rm diff}^{\gamma^*p}/dM_{X}$
extracted with the $M_X^2$ method
for each ($Q^2$, $W$, $M_X$) bin
are given in Table~\ref{tab:xsection}.
The subtraction of the double-dissociative background entails a
$\pm$21\% uncertainty in the
normalisation of the cross sections obtained with the $M_X^2$
method. 

\section{Results and discussion on the $\protect\boldmath{W}$ dependence 
of the diffractive and total cross sections}

The energy dependence of the photon-dissociative cross sections can be 
successfully described by a power of $W$, both for photoproduction
\cite{h1mx,zeusphp} and for DIS \cite{h1diff,zeusdiff94} -- at least in
the region of small $x_{\pom}$ values where the exchange of subleading
Regge trajectories can be neglected. Although the
experimental uncertainties are  large, the value of this 
power is different for the two regimes.
This is  analogous to the behaviour observed for the $W$
dependence
of the virtual photon-proton total cross section,
$\sigma^{\gamma^*p}_{\rm tot}$ \cite{zeusbpt,zeuspheno}:
the slow rise of $\sigma^{\gamma^*p}_{\rm tot}$ at high $W$ observed
in photoproduction becomes faster at high $Q^2$; the transition
takes place for $Q^2 \sim 1$~GeV$^2$.
In this section, the $W$ dependence of the photon-dissociative cross section, 
$d\sigma_{\rm diff}^{\gamma^*p} / dM_X$,  is studied in this transition
region and is 
compared to the $W$ dependence of $\sigma^{\gamma^*p}_{\rm tot}$ by 
considering the ratio of $d\sigma_{\rm diff}^{\gamma^*p}/dM_X$ to 
$\sigma^{\gamma^*p}_{\rm tot}$.

\subsection{The $\protect\boldmath{W}$ dependence of 
        the diffractive cross section}
\label{wdependence}

\begin{figure} \begin{center}
\includegraphics[clip,scale=0.80]{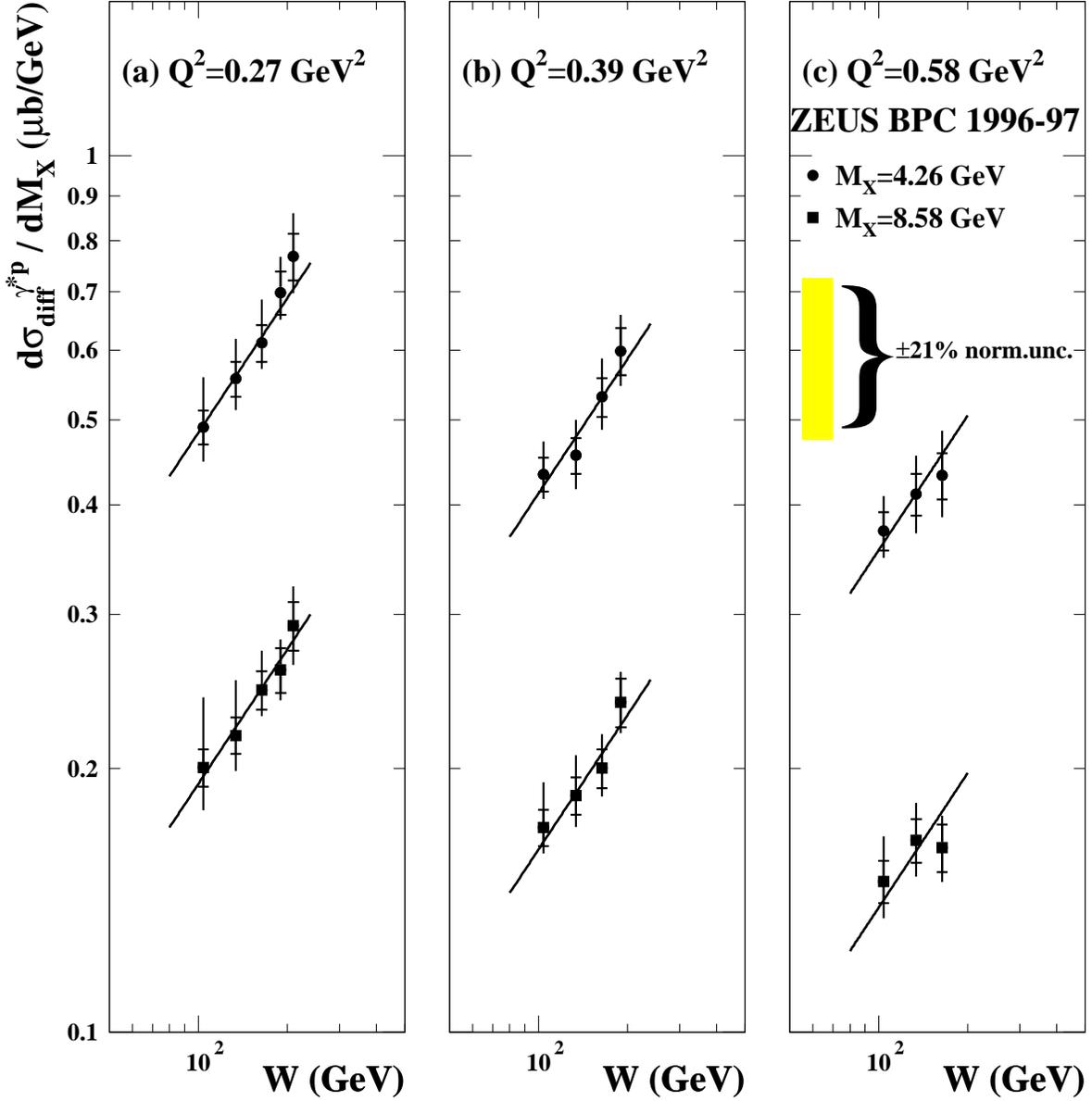} \end{center}
\caption{
        Diffractive cross sections for (a) $Q^2=0.27$~{\rm GeV}$^2$, (b)
        $Q^2=0.39$~{\rm GeV}$^2$, and (c) $Q^2=0.58$~{\rm GeV}$^2$
        for two different $M_{X}$
         ranges as a function of $W$. 
         The inner bars indicate the size
         of the statistical uncertainties; the outer bars show the size of the 
         statistical and systematic uncertainties added in quadrature. 
        The points were corrected for the double-dissociative background; 
        the associated $\pm21$\% normalisation uncertainty is not included,
        but is shown separately as a shaded error band. 
The normalisation uncertainty associated with the luminosity measurement is not
shown. 
         The lines show the results of the fit described in the text.
} \label{fig:xsection_masahide} \end{figure}

\begin{figure}
\begin{center}
\includegraphics[clip,scale=0.80]{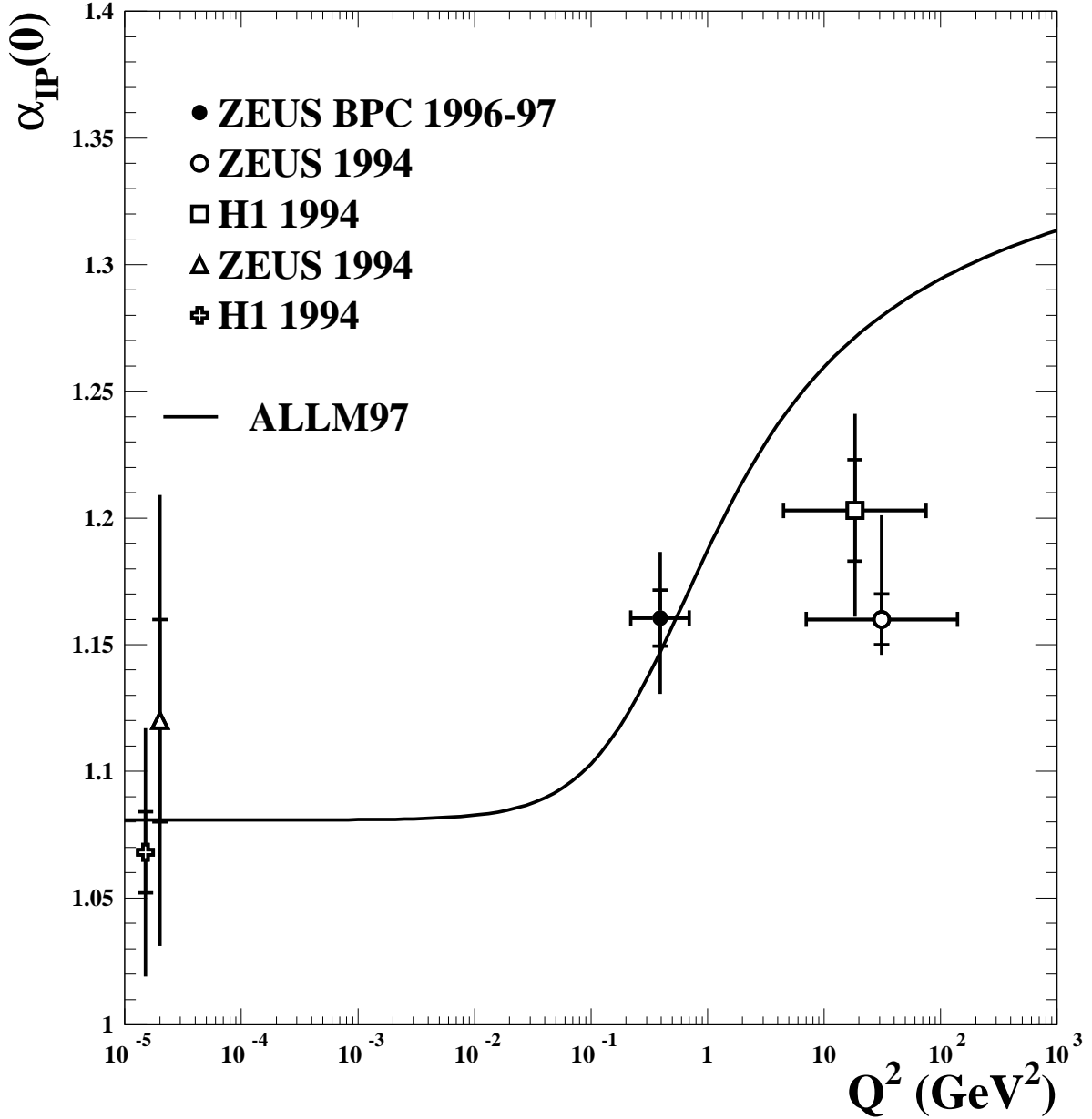}
\end{center}
\caption{Results for $\alpha_{I\hspace{-0.1cm}P}(0)$
         in different $Q^2$ regions.
         The value of $\alpha_{I\hspace{-0.1cm}P}(0)$
         obtained from this analysis is shown as the solid circle.
         The open symbols show the results 
         from the photoproduction \cite{h1mx,zeusphp}
         and DIS diffractive analyses \protect\cite{h1diff,zeusdiff94}.
         The inner  bars indicate the size
         of the statistical uncertainties; the outer bars show the size of the 
         statistical and systematic uncertainties added in quadrature.
   The line is 
   from the ALLM97 parameterisation~\protect\cite{allm97} of
the $\gamma^*p$ total cross-section data.}
         \label{fig:allm}
\end{figure}

Figure \ref{fig:xsection_masahide} shows
the values of the diffractive cross sections
extracted with the $M_X^2$ method in the BPC region
as a function of $W$ for three $Q^2$ and two $M_X$ bins. 
The form
\begin{equation}
        \frac{d\sigma_{\rm diff}^{\gamma^*p}}{dM_{X}}
        =A_i \cdot W^{a_{\rm diff}}
        \label{cross-apom4}
\end{equation}
was fitted to these data, where $a_{\rm diff}$ is a global parameter and
the normalisation parameters $A_i$
were left free to vary for each ($Q^2$,$M_X$) bin.
The results of the fit, taking into account the  correlations between
adjacent bins, are shown in Fig.~\ref{fig:xsection_masahide}; they
give a good description of the data ($\chi^2/ndf = 0.51$, calculated using 
the statistical uncertainties only).
The fitted value of the power of $W$ is
\begin{equation}
        a_{\rm diff} =
        0.510 \pm 0.043 \mbox{(stat.)}^{+0.102}_{-0.122} \mbox{(syst.)}.
        \nonumber
\end{equation}
Expressing the $W$ dependence of the cross section in terms of an
effective Pomeron intercept~\cite{pdbc}, $\bar{\alpha}_{I\hspace{-0.1cm}P}$,
as
\begin{equation}
        \frac{d\sigma_{\rm diff}^{\gamma^*p}}{dM_{X}}\propto
        (W^{2})^{2\bar{\alpha}_{I\hspace{-0.1cm}P}-2}, \nonumber
        \label{cross-apom2}
\end{equation}
\noindent
the fitted value of $a_{\rm diff}$ corresponds to
\begin{equation}
        \bar{\alpha}_{\pom} =
        1.128\pm0.011 \mbox{(stat.)}^{+0.026}_{-0.030} \mbox{(syst.)}.
        \nonumber
        \label{effapom}
\end{equation}
This value of $\bar{\alpha}_{I\hspace{-0.1cm}P}$ 
can, in turn, be related to the Pomeron intercept,
$\alpha_{I\hspace{-0.1cm}P}(0)$, via
\begin{displaymath}
        \bar{\alpha}_{I\hspace{-0.1cm}P} =
        \alpha_{I\hspace{-0.1cm}P}(0) -
        \alpha_{I\hspace{-0.1cm}P}'\cdot |\bar{t}|,
        \label{relation}
\end{displaymath}
where $|\bar{t}|$ is the mean value of $|t|$.
The value of $\alpha_{I\hspace{-0.1cm}P}(0)$,
obtained assuming $\alpha_{I\hspace{-0.1cm}P}'=0.25~$GeV$^{-2}$ 
and using $|\bar{t}|=0.13\,$GeV$^{2}$ \cite{lps_f2d3, tslope_php},
is  $\alpha_{\pom}(0)=1.161 \pm0.011 \mbox{(stat.)}^{+0.026}_{-0.030} 
\mbox{(syst.)}$; it
is shown in Fig.~\ref{fig:allm} together with the values determined from
photoproduction and from higher-$Q^2$ 
measurements~\cite{h1mx,zeusphp,h1diff,zeusdiff94}. The quoted systematic uncertainty
does not include the uncertainty on $\alpha_{\pom}^{\prime}$, which was
also not
included in the other results presented in Fig.~\ref{fig:allm}. 
The value of
$\alpha_{I\hspace{-0.1cm}P}(0)$ from the present measurement at low
$Q^2$ does not differ significantly from the values at higher $Q^2$.

Equation~(\ref{cross-apom4}) was also fitted to the data allowing
different values 
for the parameter $a_{\rm diff}$ in the three $Q^2$ bins of the measurement;
the three resulting values of $a_{\rm diff}$  are  compatible 
with the global value. 
The data used in the fit have values of 
$x_{\pom}$ typically much smaller than 0.01, with the exception of the bin
with lowest $W$ and highest $M_X$ values, which receives contributions
from
$x_{\pom}$ values up to $x_{\pom}=0.018$. It was assumed that Pomeron
exchange dominates in this region, and no attempt was made to include
secondary Reggeon exchange in the fit.
Finally, it should be noted that a possible $W$ dependence of the
double-dissociative fraction would affect the extracted value of
$\alpha_{\pom}$.

Figure~\ref{fig:allm} also shows $\alpha_{I\hspace{-0.1cm}P}(0)$ as
obtained from the ALLM97 parameterisation~\cite{allm97} of the $\gamma^*p$
total cross section, which gives a good representation of the
inclusive $F_2$ data for the entire $Q^2$ range.
The value of $\alpha_{\pom}(0)$ from ALLM97 is consistent with the present 
determination from the diffractive data in the BPC region, whereas in the
DIS region it
is higher than the H1 and ZEUS diffractive
measurements~\cite{h1diff,zeusdiff94}.

\begin{figure}
\begin{center}
\includegraphics[clip,scale=0.80]{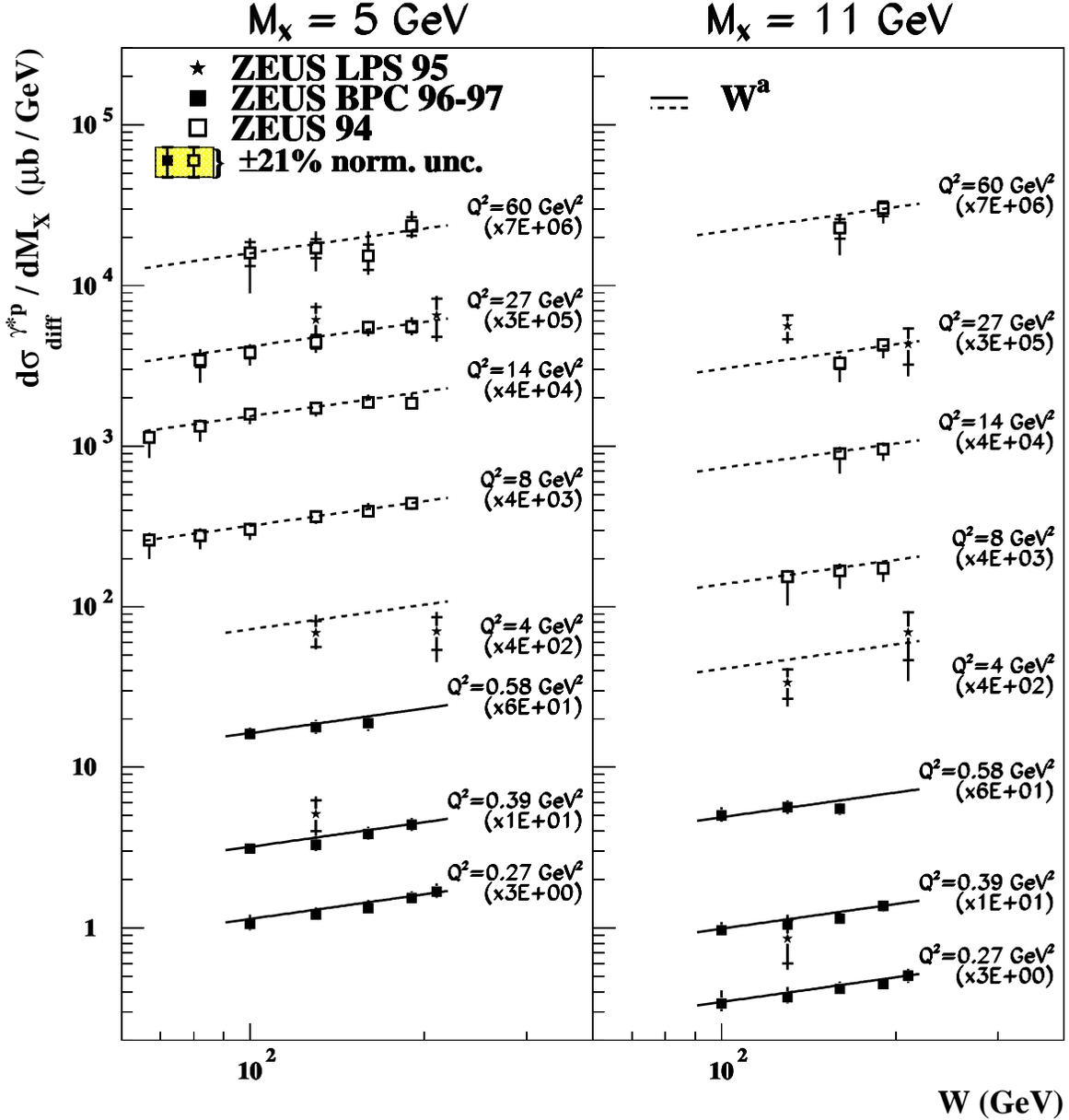}
\end{center}
\vspace{-1cm}
\caption{
        Diffractive cross sections for different $Q^2$ and $M_{X}$
        values as a function of $W$. 
        The results obtained with the LPS method are shown as stars.
        The inner  bars indicate the size
        of the statistical uncertainties; the outer bars show the size of the 
        statistical and systematic uncertainties added in quadrature. 
        The low-$Q^2$ points obtained with the $M_X^2$ method
        (full squares) were corrected for the double-dissociative background; 
        the corresponding $\pm21$\% normalisation uncertainty is not included,
        but is shown separately as a shaded band.
        The normalisation uncertainty associated with the luminosity
        measurement is not shown. 
        The open squares at high $Q^2$ are from a previous ZEUS
        publication~\protect\cite{zeusdiff94} and have been corrected for the
        double-dissociative background using the same estimate 
        as for the low-$Q^2$
        points, as discussed in the text.  
The solid lines are the results of the fit to the BPC data
described in
the text, which also gives a good representation of the higher-$Q^2$ 
data (dashed lines).
}
\label{fig:xsection}
\end{figure}

The LPS cross sections are presented in Fig.~\ref{fig:xsection}; 
they are in agreement with the previous 
ZEUS measurements at large $Q^2$ and with the present
BPC data obtained with the $M_X^2$ method. The previous ZEUS 
data~\cite{zeusdiff94} obtained by the $M_X^2$ method have also been corrected
for the residual double-dissociative background using the value
of $R_{M_X}$ given in Section~\ref{pdiss}; 
to make a direct comparison with the earlier data, the BPC cross sections 
from 
Table~\ref{tab:xsection} have been interpolated to $M_X=5$ and 11~GeV
using bin-centring corrections based on EPSOFT.
The solid lines in Fig.~\ref{fig:xsection} correspond to the fit of 
Eq.~(\ref{cross-apom4}) to the BPC data alone, which also provides a
 good description of the DIS data (dashed lines).
 Figures~\ref{fig:allm} and~\ref{fig:xsection} thus
show that the $W$ dependence of the inclusive diffractive cross section
exhibits no significant changes from the BPC to the DIS region.

\subsection{Comparison of the $\protect\boldmath{W}$ dependence of 
        the diffractive and the total cross sections
}

\begin{figure}
\begin{center}
\includegraphics[clip,scale=0.80]{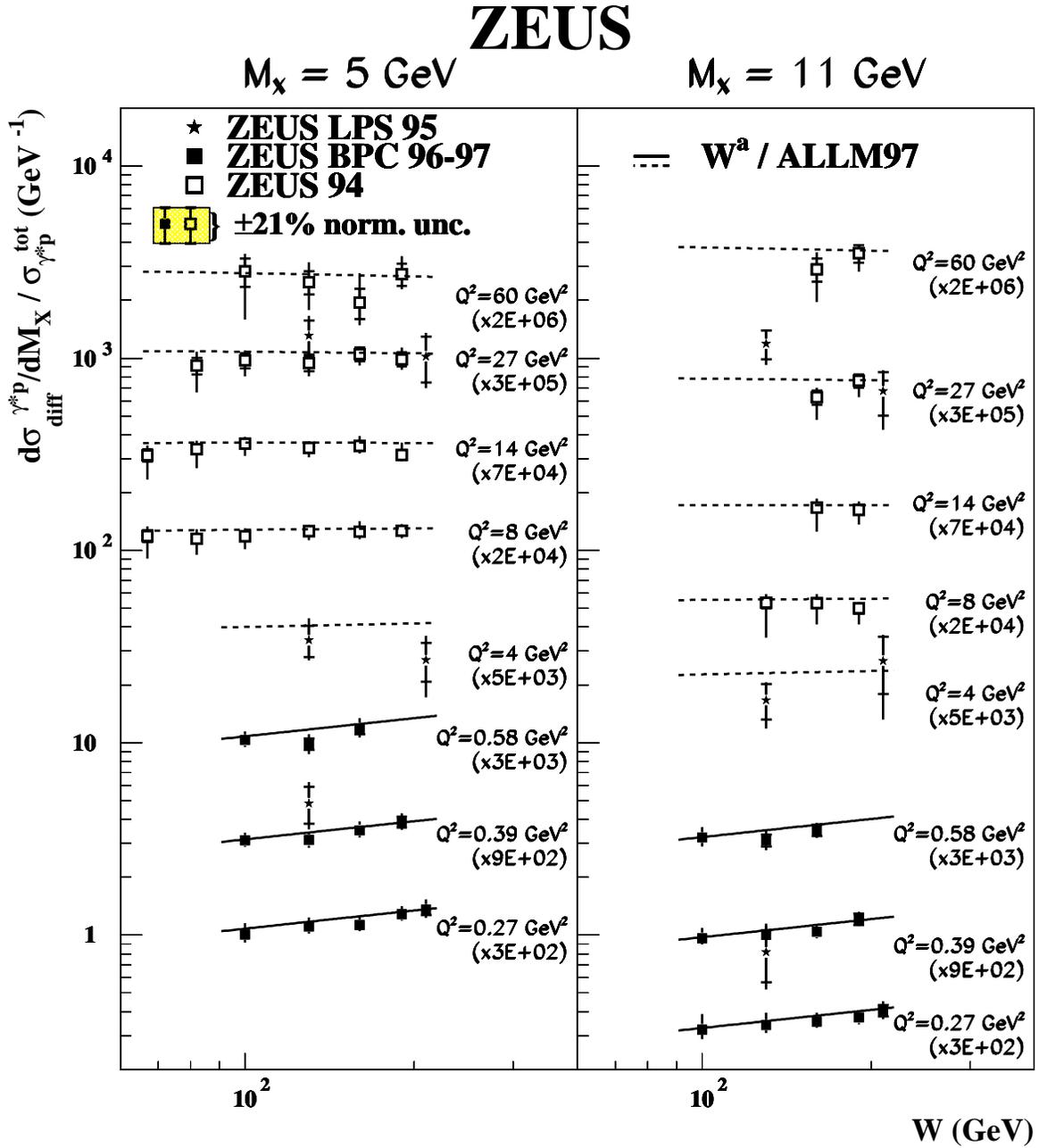}
\end{center}
\caption{
        The ratio of the diffractive cross section
        to the $\gamma^*p$ total cross section
        for different $Q^2$ and $M_X$ values 
        as a function of $W$. Other details are as in the caption to 
        Fig.~\ref{fig:xsection}. The lines denote the fit shown
        in Fig.~\ref{fig:xsection}
        divided by the corresponding values of
        $\sigma^{\gamma^*p}_{\rm tot}$ from the ALLM97 parameterisation.
}
\label{fig:xsection_ratio_w}
\end{figure}

The $W$ dependences of the diffractive and total cross sections
were directly compared by studying their ratio
\begin{equation}
r=
\frac{d\sigma_{\rm diff}^{\gamma^*p}/dM_X}
     {\sigma ^{\gamma^*p}_{\rm tot}}. \nonumber
\label{wratio}
\end{equation}
This ratio is plotted as a function of $W$
in Fig.~\ref{fig:xsection_ratio_w}, where 
the values of the diffractive cross sections shown in
Fig.~\ref{fig:xsection} were divided by the corresponding
values of the $\gamma^*p$ total cross section,
$\sigma^{\gamma^*p}_{\rm tot}$ \cite{zeusbpt,F29697}.
The  lines denote the fit shown in Fig.~\ref{fig:xsection}
divided by the corresponding values of the ALLM97 parameterisation for
$\sigma^{\gamma^*p}_{\rm tot}$.
The lines give a good representation of all the data.
While there is a clear increase in $r$ as a function of $W$ for $Q^2 < 1$
GeV$^2$, for higher $Q^2$ the distribution is flat in $W$.

The form $ r = N_i \cdot W^{\rho} $
was fitted to the BPC data measured with the $M_X^2$ method; here,
$\rho$ is a global parameter and the normalisation parameters, 
$N_i$, were left free to vary for each ($Q^2, M_X$) bin.
The fit gives a good description (not shown) of the data with
$\rho=0.24\pm 0.07$, where the uncertainty is derived from the fit, consistent
with the expectation~\cite{zeusdiff94} from Regge theory that

\begin{equation}
r= \frac {(d\sigma_{\rm diff}^{\gamma^*p}/dM_X)}
{\sigma ^{\gamma^*p}_{\rm tot}} \propto \frac {(W^2)^{2\bar{\alpha}_{\pom}-2}}
{(W^2)^{\alpha_{\pom}(0)-1}} =
W^{2(2\bar{\alpha}_{\pom}-\alpha_{I\hspace{-0.1cm}P}(0)-1)} \approx
W^{0.19}. \nonumber
\end{equation}

\noindent
This result suggests a different behaviour from that found
in the DIS region, where the value $\rho=0.00 \pm
0.03$~\cite{zeusdiff94} indicates that the diffractive and inclusive
cross sections have the same $W$ dependence, contrary to the expectations 
of Regge theory.

In summary, in the BPC region the $W$ dependence of the diffractive cross
section is compatible with that expected from Regge phenomenology.
The ratio between diffractive and total cross sections
grows with $W$ at a rate consistent with Regge theory.
This is in contrast to the DIS region, 
where the expectations of Regge theory for the ratio of diffractive and
total cross sections are not fulfilled, since the ratio is flat as a
function of $W$. This difference in the $W$ dependence of the ratio is
reflected in the fact that the values of $\alpha_{\pom}(0)$ extracted from 
the diffractive cross section and from $\sigma_{\rm tot}^{\gamma^*p}$ 
are similar in the BPC region, but not in the DIS region.

\section{
        \protect{Results and discussion on the \boldmath{$Q^2$}} 
dependence of diffractive and total cross sections}

The $Q^2$ dependence of $\sigma^{\gamma^*p}_{\rm tot}$
has been observed to change around $Q^2 \sim 1$~GeV$^2$~\cite{zeusbpt}:
compared to the approximate $1/Q^2$ scaling behaviour found at high $Q^2$,
data at $Q^2~ \lsim~1\,$GeV$^2$ exhibit a weaker $Q^2$ dependence, 
with $\sigma^{\gamma^*p}_{\rm tot}$ being nearly independent of $Q^2$
at the lowest $Q^2$ values measured. This is consistent with the
expectation from the conservation of the electromagnetic current that
$\sigma^{\gamma^*p}_{\rm tot}$ approaches a constant or, equivalently, 
that $F_2$ vanishes like $Q^2$ as $Q^2 \rightarrow 0$.

In this section, the $Q^2$ dependence of
the diffractive cross section, $d\sigma_{\rm diff}^{\gamma^*p}/dM_X$,
is studied and the question is addressed of
whether and where a transition similar to that observed for
$\sigma^{\gamma^*p}_{\rm tot}$ occurs for the diffractive
dissociation of virtual photons.

\subsection{
        The \protect{\boldmath{$Q^2$}} dependence of
        the diffractive cross section
}

\begin{figure}
        \begin{center}
        \includegraphics[clip,scale=0.80]{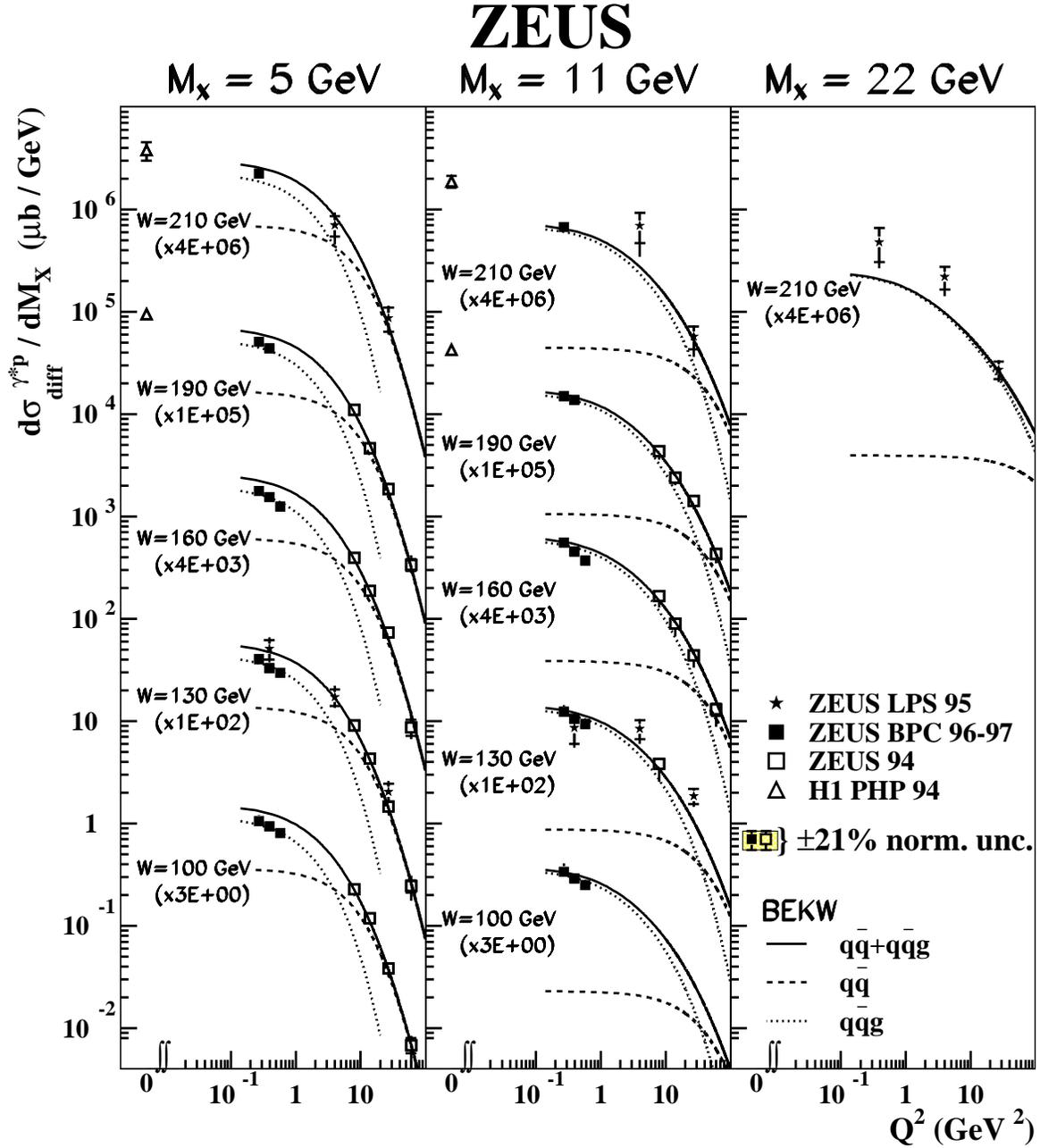}
        \end{center}
        \caption{
                The values of $d\sigma_{\rm{diff}}^{\gamma^*p}/dM_X$
                for different $W$ and $M_X$ values as a function of $Q^2$.
                Other details are as given in the caption to
                Fig.~\ref{fig:xsection}.
                The solid lines are the results of the BEKW parameterisation
                described in the text;
                the dotted (dashed) lines are the results of the same
                parameterisation for the $q\bar{q}g$ 
                ($q\bar{q}$) contribution alone.
            Note the break in the $Q^2$ scale below $\sim 10^{-2}$ GeV$^2$. 
        }
        \label{dsigma_dmx_vs_q2}
\end{figure}

Figure~\ref{dsigma_dmx_vs_q2}
shows the diffractive cross sections,
$d\sigma_{\rm diff}^{\gamma^*p}/dM_X$,
as a function of $Q^2$ in bins of $W$ and $M_X$.
The present measurements are plotted together with
previous ZEUS results~\cite{zeusdiff94}, obtained
with the $M_X^2$ method in the DIS region, and H1 results~\cite{h1php},
obtained with the rapidity-gap method in photoproduction for $M_N <$ 1.6
GeV and $|t|<1$~GeV$^2$; bin-centring corrections based on EPSOFT,
analogous to those
described in Section~\ref{wdependence}, were applied,
where necessary, to both ZEUS and H1 results. No further corrections
were applied to the H1 data; notably, no attempt was made to correct for 
the double-dissociative background.

In Fig.~\ref{dsigma_dmx_vs_q2},
a change in the $Q^2$ dependence of $d\sigma_{\rm diff}^{\gamma^*p}/dM_X$
as $Q^2$ increases is apparent and is similar to that observed
in the $\sigma^{\gamma^*p}_{\rm tot}$ data: at low
$Q^2$, the data do not exhibit a strong $Q^2$ dependence, while 
at larger $Q^2$,
the cross section falls rapidly for increasing $Q^2$.
Figure~\ref{f2d_vs_q2} shows $\xpom F_2^{D(3)}$
as a function of $Q^2$ for fixed $W$ and $M_X$;
while at large $Q^2$ the data do not exhibit a strong $Q^2$
dependence, $\xpom F_2^{D(3)}$ falls by a factor of about ten between 
$Q^2 \approx 8$~GeV$^2$ and $Q^2 \approx 0.2$~GeV$^2$. 

\subsection{Discussion
}

The diffractive dissociation of the virtual photon can  be 
described by perturbative QCD (pQCD) since the photon's virtuality, $Q^2$,
provides a hard scale. In particular, in the proton rest frame,
the reaction can be viewed as a sequence of three successive 
processes~\cite{review_halina,reviews}:
the photon fluctuates into a ${q\bar{q}}$ (or ${q\bar{q}g}$) state, 
the ${q\bar{q}}$ dipole scatters off the proton target and, finally,  
the scattered ${q\bar{q}}$  pair produces the final state.
At high centre-of-mass energies of the $\gamma p$ system, these processes are
widely separated in time. The ${q\bar{q}}$ fluctuation is described in terms of
the photon wave-function derived from QCD.  
The interaction of the ${q\bar{q}}$ dipole with the proton is mediated, in 
lowest order, by the exchange of two gluons in a colour-singlet state.

The present results have been compared to the model of Bartels 
{\it et al.} (BEKW)~\cite{bekw}. In this model,
neglecting the contribution of longitudinally polarised photons,
the dominant (leading-twist) contributions to the diffractive
structure function in the kinematic domain
of the present measurement come from the fluctuations
of the photon into either a ${q\bar{q}}$ pair ($F_{q\bar{q}}^{T}$)
or a ${q\bar{q}g}$ state ($F_{q\bar{q}g}^{T}$).
The $\beta$ spectra of these two components
are determined by rather general properties of
the photon wave-function:
$F_{q\bar{q}}^T$ behaves like $\beta\,(1-\beta)$ and
$F_{q\bar{q}g}^T$ like $(1-\beta)^{\gamma}$, where 
$\gamma$ = 3.9~\cite{zeusdiff94,bekw}.
At large $\beta$, $q\bar{q}$ production dominates over $q\bar{q}g$ production,
while, at small $\beta$, $q\bar{q}g$ production becomes dominant.
$F_{q\bar{q}}^T$ has no $Q^2$ dependence;
$F_{q\bar{q}g}^T$  is of order
$\alpha_S$ and has a logarithmic $Q^2$ dependence of the type
$\log{(1+Q^2/Q^2_0)}$, where
the scale parameter $Q^2_0$ is taken to be 1~GeV$^2$.
The model does not fix the $x_{I\hspace{-0.1cm}P}$ dependence of
$F_{q\bar{q}}^T$  and $F_{q\bar{q}g}^T$, but assumes for both a power-like
behaviour $x_{I\hspace{-0.1cm}P}^{-n_{\rm diff}(Q^2)}$, where the exponent
$n_{\rm diff}$ is determined from fits to the data.

\begin{figure}
        \begin{center}
                \includegraphics[clip,scale=0.80]{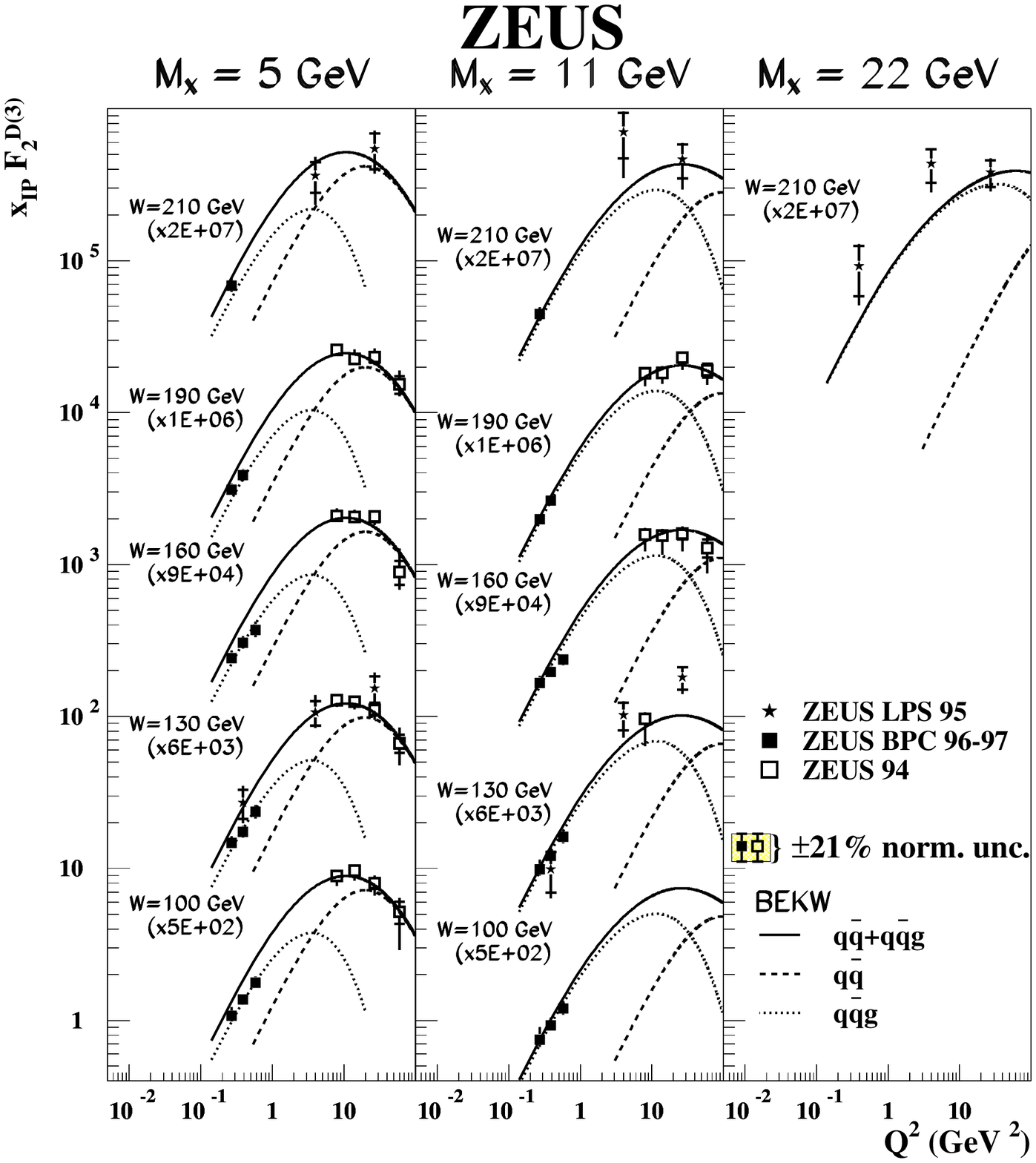}
        \end{center}
        \caption{
                The values of $x_{I\hspace{-0.1cm}P}F_2^{D(3)}$
                for different $W$ and $M_X$ values as a function of $Q^2$.
                Other details are as given in the caption to Fig.~\ref{fig:xsection}.
                The solid lines are the results of the BEKW parameterisation
                described in the text;
                the dotted (dashed) lines are the results of the same
                parameterisation for the $q\bar{q}g$ 
                ($q\bar{q}$) contribution alone.
        }
        \label{f2d_vs_q2}
\end{figure}

A comparison of the BEKW parameterisation with the present data is
shown in Figs.~\ref{dsigma_dmx_vs_q2} and \ref{f2d_vs_q2}.
The values of the parameters, including the normalisation of
the $F_{q\bar{q}}^T$ and $F_{q\bar{q}g}^T$ components, were taken
from a fit
to the previous ZEUS results~\cite{zeusdiff94}, with the exception of the
$x_{I\hspace{-0.1cm}P}$ exponent, for which a constant value
corresponding to $a_{\rm diff}/2$, determined from Eq.~(\ref{cross-apom4}),
was used. The DIS data at high $Q^2$
constrain the parameterisation of the $\beta$ dependence
of $F_{q\bar{q}}^{T}$ (dashed lines)
at low $M_X$ and of $F_{q\bar{q}g}^{T}$ (dotted lines) at high $M_X$.
The logarithmic $Q^2$ dependence of $F_{q\bar{q}g}^{T}$
is probed only in the highest-$Q^2$ region and is less well constrained.

The $Q^2$ dependence of $F_{q\bar{q}g}^{T}$
becomes crucial in the transition to low $Q^2$.
In fact, as $Q^2$ decreases from the DIS into the BPC region,
for a given value of $M_X$, $\beta$ also decreases:
the BPC data in Figs.~\ref{dsigma_dmx_vs_q2} and \ref{f2d_vs_q2}
correspond to values of $\beta$ that are typically two orders
of magnitude smaller than those in the DIS data and thus, in the BPC region,
the contribution from the fluctuation of the
photon into a $q\bar{q}g$ system becomes dominant.
  While extrapolating the BEKW
        parameterisation to low $Q^2$ may not be formally justified,
        it is interesting to note that, in $F_{q\bar{q}g}^{T}$,
        conservation of the electromagnetic current is assured by the fact 
that $\log{(1+Q^2/Q^2_0)}$ vanishes like $Q^2/Q^2_0$ as $Q^2 \rightarrow 0$.
        The transition from the linear behaviour at low $Q^2$
        to the logarithmic behaviour at higher $Q^2$
        is controlled by the scale parameter $Q^2_0$;
        the choice $Q^2_0 = 1$~GeV$^2$ successfully describes the BPC
data.

\subsection{
        Comparison of the \protect\boldmath{$Q^2$} dependence of 
        the diffractive and the total cross sections
}

\begin{figure}
        \begin{center}
        \includegraphics[clip,scale=0.80]{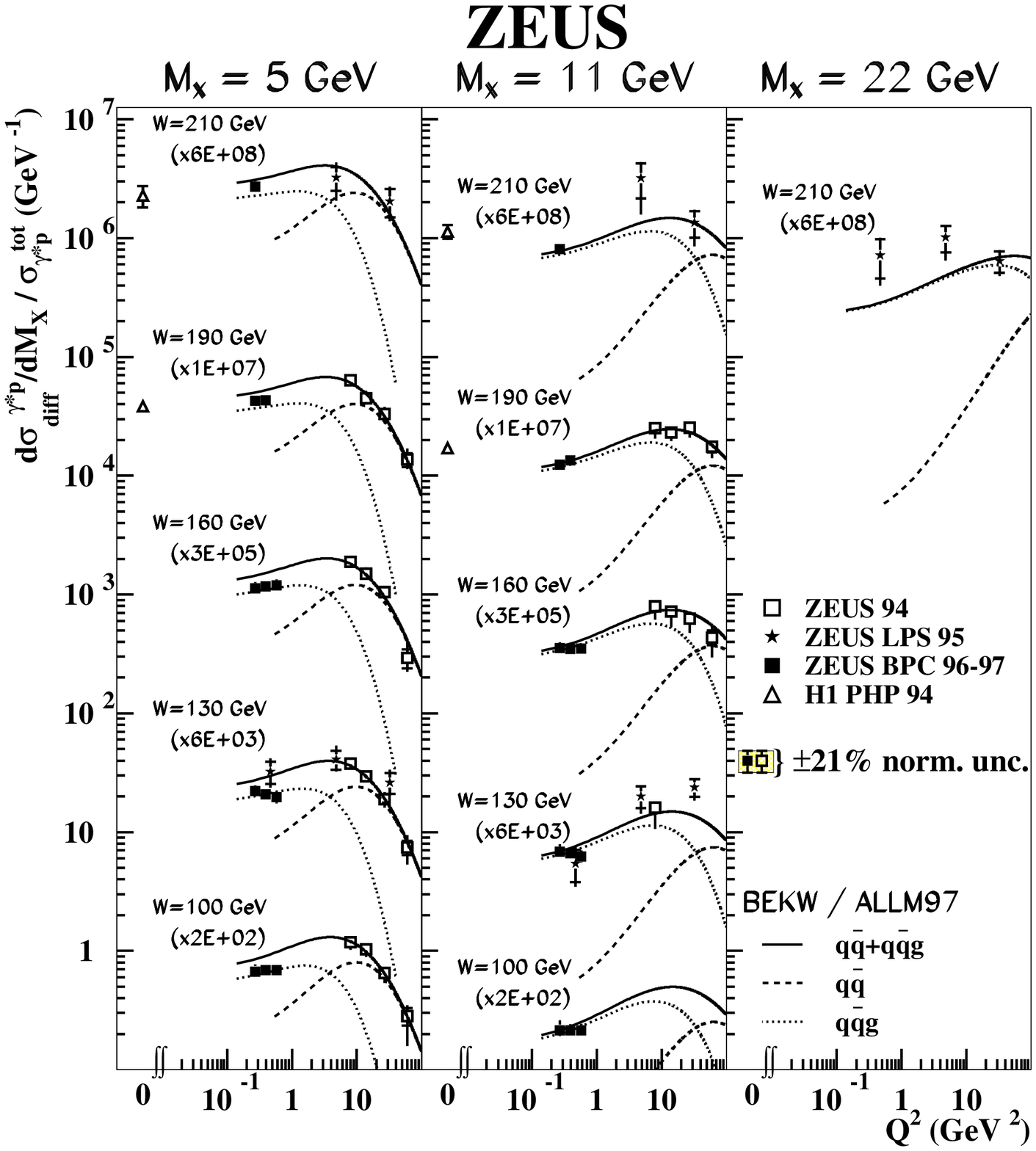}
        \end{center}
\vspace{-1cm}
        \caption{
        The ratio of the diffractive cross section
        to the $\gamma^*p$ total cross section
        for different $W$ and $M_X$ values as a function of $Q^2$.
      Other details are as given in the caption to Fig.~\ref{fig:xsection}.
        The solid lines are the results of the BEKW parameterisation
        described in the text,
        divided by the corresponding
        $\sigma^{\gamma^*p}_{\rm tot}$ values from
        the ALLM97 parameterisation;
        the dotted (dashed) lines are the results of the same
        parameterisations for the $q\bar{q}g$ 
        ($q\bar{q}$) contribution alone.
            Note the break in the $Q^2$ scale below $\sim 10^{-2}$ GeV$^2$. 
        }
        \label{dsigma_dmx_vs_q2_ratio}
\end{figure}

Figure~\ref{dsigma_dmx_vs_q2_ratio} shows  the ratio 
$r= (d\sigma_{\rm diff}^{\gamma^*p}/dM_X)/
\sigma^{\gamma^*p}_{\rm tot}$ as a function of $Q^2$ for different
$W$ and 
$M_X$ bins. At low $Q^2$, the $Q^2$ dependence of the diffractive cross 
section is similar to that of $\sigma^{\gamma^*p}_{\rm tot}$. 
In the DIS regime, $d\sigma_{\rm diff}^{\gamma^*p}/dM_X$
decreases with $Q^2$ more rapidly than $\sigma^{\gamma^*p}_{\rm tot}$.
This is more evident for small values of $M_X$.
In addition, the ratio 
$r$ appears to increase between the BPC and the DIS 
regions.

Also shown in the figure are the results of the BEKW
parameterisation of the diffractive cross-section
$d\sigma_{\rm diff}^{\gamma^*p}/dM_X$, shown in
Fig.~\ref{dsigma_dmx_vs_q2},
divided by the values of
$\sigma^{\gamma^*p}_{\rm tot}$
given by the ALLM97 parameterisation~\cite{allm97}.
There is reasonable agreement between these parameterisations
and the data, indicating that the data
may be qualitatively described by an appropriate choice of the
relative fractions of the $q\bar{q}$ and $q\bar{q}g$ contributions.

\section{Summary}

The diffractive dissociation of virtual photons, 
$\gamma^* p \rightarrow X p$, has been studied at HERA 
at low $Q^2$ ($0.17<Q^2<0.70$~GeV$^2$)
and in deep inelastic scattering (DIS) ($3<Q^2 < 80$~GeV$^2$).
The diffractive signal has been selected either by 
requiring the detection of 
a final-state proton with at least 97\% of the incoming proton-beam
energy, or by exploiting
the different properties of the $M_X$ distributions
for diffractive and non-diffractive events.

The $W$ dependence of the low-$Q^2$ cross-section data obtained with the
$M_X^2$ method ($3<M_X<12.2$~GeV) has been found to be compatible
with a single power of $W$,
which corresponds to a Pomeron intercept of $\alpha_{\pom}(0) =
1.161\pm0.011 \mbox{(stat.)}^{+0.026}_{-0.030} \mbox{(syst.)}$.
This is consistent with that previously observed in the DIS
regime.
Thus, the significant change in the $W$ dependence exhibited by the
$\gamma{^*}p$ total cross section in the transition from low $Q^2$ to DIS
is not visible in the diffractive cross section.
To elucidate this difference, the $W$ dependence of the ratio, $r$, of the
diffractive cross section to the $\gamma^*p$ total cross section was studied
at low $Q^2$ and was found to rise with $W$, $r\propto W^{0.24 \pm 
0.07}$, in agreement with the expectation
from Regge theory. This is in contrast to the observation at higher $Q^2$
that this ratio is independent of $W$.

The $Q^2$ dependence of the diffractive cross section changes as $Q^2$
increases up to the DIS regime: 
while at low $Q^2$ the data do not exhibit a strong
$Q^2$ dependence, at larger $Q^2$ the cross section falls rapidly
for increasing $Q^2$. This change of behaviour occurs for values of
$Q^2$ around 1 GeV$^2$ and is analogous to that observed in the total 
$\gamma{^*}p$ cross section.
The ratio of the diffractive cross section to the $\gamma^*p$ total cross
section was studied as a function of $Q^2$.
At low $Q^2$, the ratio $r$ shows little dependence on $Q^2$, indicating
that the $Q^2$ dependence of the diffractive cross
section is similar to that of $\sigma^{\gamma^*p}_{\rm tot}$.
The ratio increases between the BPC and the DIS regions.
In the DIS regime for low $M_X$, the ratio decreases with increasing $Q^2$,
indicating that the diffractive cross section
decreases with $Q^2$ more rapidly than the $\gamma^*p$ total cross section.

The main features of the data,
reproduced
by a parameterisation based on the BEKW model,
indicate that the framework
in which the incoming virtual photon
fluctuates into a quark-antiquark pair is, in general, adequate
to describe diffractive processes in $ep$
collisions from the BPC to the DIS region.
At the same time, the data suggest the increasing importance of 
the contribution from $q\bar{q}g$ states at low $Q^2$.
It is interesting that 
 the ratio of the diffractive cross section to the total cross section
shows a change from a $W^{0.24\pm 0.07}$ dependence for $Q^2<0.7$ GeV$^2$
to $W^{0.00\pm 0.03}$ for $Q^2>3$ GeV$^2$.
This complex behaviour of diffraction as a function of both $Q^2$ and $W$
reveals a rich testing ground for future theoretical models.

\section*{Acknowledgements}

We thank the DESY directorate for their encouragement, and acknowledge 
the support of the DESY computing and network services. We are especially 
grateful to the HERA machine group: collaboration with them was crucial to 
the successful installation and operation of the leading proton
spectrometer.
The design, construction  and installation of the ZEUS
detector have been made possible by the ingenuity and efforts of many people 
from DESY and home institutes who are not listed as authors. Finally, 
it is a pleasure to thank J. Bartels, 
K. Golec-Biernat, N.N. Nikolaev, M.G. Ryskin and
M. Strikman for many useful discussions.

\newpage

\renewcommand{\arraystretch}{1.3}
\begin{table}
\begin{center}
\begin{tabular}{|c|c|c||c|c|c||c|c|c||c|}
\hline
$Q^{2}_{\rm min}$ & $Q^{2}_{\rm max}$ & $ Q^{2}$ &
$W_{\rm min}$ & $W_{\rm max}$ & $ W $ &
$M_{X, \rm min}$ & $M_{X,\rm max}$ & $ M_X $ &
{$\frac{d^{2}\sigma_{\rm diff}^{\gamma^*p}}{dM_X dt}$} \\
 \hline
\multicolumn{3}{ |c||}{(GeV$^{2}$)} &
\multicolumn{3}{ c||}{(GeV)} &
\multicolumn{3}{ c||}{(GeV)} &
($\mu$b/GeV$^3$) \\
 \hline
 \hline
  0.17 & 0.70 & 0.39 &  90 & 250 & 130 & 3.00 &  6.05 &  5 & $  0.867\pm  0.186^{+ 0.170}_{-  0.139}$ \\
   &  &  &  90 & 250 & 130 & 6.05 & 12.20 & 11 & $  0.144\pm  0.043^{+  0.040}_{-  0.028}$ \\
   &  &  & 165 & 250 & 210 &12.20 & 38.00 & 22 & $  0.202\pm  0.074^{+  0.034}_{-  0.052}$ \\
 \hline
 \hline
  3 & 9 & 4 &  80 & 165 & 130 & 3.00 &  6.05 &  5 & $  0.346\pm  0.063^{+  0.081}_{-  0.039}$ \\
   &  &  & 165 & 250 & 210 & 3.00 &  6.05 &  5 & $  0.349\pm  0.080^{+  0.083}_{-  0.095}$ \\
   &  &  &  80 & 165 & 130 & 6.05 & 12.20 & 11 & $  0.172\pm  0.036^{+  0.019}_{-  0.035}$ \\
   &  &  & 165 & 250 & 210 & 6.05 & 12.20 & 11 & $  0.350\pm  0.115^{+  0.050}_{-  0.133}$ \\
   &  &  & 165 & 250 & 210 &12.20 & 33.00 & 22 & $  0.098\pm  0.024^{+  0.006}_{-  0.025}$ \\
 \hline
 \hline
  9 & 80 & 27 &  80 & 165 & 130 & 3.00 &  6.05 &  5 & $  0.042\pm  0.008^{+  0.008}_{-  0.005}$ \\
   &  &  & 165 & 250 & 210 & 3.00 &  6.05 &  5 & $  0.044\pm  0.012^{+  0.008}_{-  0.007}$ \\
   &  &  &  80 & 165 & 130 & 6.05 & 12.20 & 11 & $  0.038\pm  0.006^{+  0.002}_{-  0.005}$ \\
   &  &  & 165 & 250 & 210 & 6.05 & 12.20 & 11 & $  0.029\pm  0.007^{+  0.003}_{-  0.008}$ \\
   &  &  & 165 & 250 & 210 &12.20 & 33.00 & 22 & $  0.014\pm  0.003^{+  0.002}_{-  0.002}$ \\
 \hline
\end{tabular}
\end{center}
\caption{
        The values of $d^2\sigma_{\rm diff}^{\gamma^*p}/dM_X dt$ 
        measured with the LPS method in the
        range $0.073 <|t|<0.40$~{\rm GeV}$^2$ with the bin ranges indicated.
     The data are at $\langle t \rangle = 0.17$~{\rm GeV}$^2$. 
     The first and second error values represent
        the statistical and systematic uncertainties, respectively.
        The normalisation
uncertainty related to the luminosity measurement is not included 
in the systematic uncertainty. 
}
\label{tbl:lps1}
\end{table}

\begin{table}
\begin{center}
\begin{tabular}{|c|c|c||c|c|c||c|c|c||c|}
\hline
$Q^{2}_{\rm min}$ & $Q^{2}_{\rm max}$ & $ Q^{2}$ &
$W_{\rm min}$ & $W_{\rm max}$ & $ W$ &
$M_{X, \rm min}$ & $M_{X,\rm max}$ & $M_X$ &
{$\frac{d\sigma_{\rm diff}^{\gamma^*p}}{dM_X}$} \\
 \hline
\multicolumn{3}{ |c||}{(GeV$^{2}$)} &
\multicolumn{3}{ c||}{(GeV)} &
\multicolumn{3}{ c||}{(GeV)} &
($\mu$b/GeV) \\
 \hline
 \hline
  0.17 & 0.70 & 0.39 &  90 & 250 & 130 & 3.00 &  6.05 &  5 & $  0.511\pm  0.110^{+ 0.100}_{-  0.082}$ \\
   &  &  &  90 & 250 & 130 & 6.05 & 12.20 & 11 & $  0.086\pm  0.026^{+  0.024}_{-  0.017}$ \\
   &  &  & 165 & 250 & 210 &12.20 & 38.00 & 22 & $  0.120\pm  0.044^{+  0.020}_{-  0.031}$ \\
 \hline
 \hline
  3 & 9 & 4 &  80 & 165 & 130 & 3.00 &  6.05 &  5 & $  0.172\pm  0.031^{+  0.040}_{-  0.019}$ \\
  &  &  & 165 & 250 & 210 & 3.00 &  6.05 &  5 & $  0.175\pm  0.040^{+  0.042}_{-  0.047}$ \\
   &  &  &  80 & 165 & 130 & 6.05 & 12.20 & 11 & $  0.084\pm  0.017^{+  0.009}_{-  0.017}$ \\
   &  &  & 165 & 250 & 210 & 6.05 & 12.20 & 11 & $  0.174\pm  0.057^{+  0.025}_{-  0.066}$ \\
   &  &  & 165 & 250 & 210 &12.20 & 33.00 & 22 & $  0.055\pm  0.014^{+  0.003}_{-  0.014}$ \\
 \hline
 \hline
  9 & 80 & 27 &  80 & 165 & 130 & 3.00 &  6.05 &  5 & $  0.020\pm  0.004^{+  0.004}_{-  0.002}$ \\
   &  &  & 165 & 250 & 210 & 3.00 &  6.05 &  5 & $  0.022\pm  0.006^{+  0.004}_{-  0.004}$ \\
   &  &  &  80 & 165 & 130 & 6.05 & 12.20 & 11 & $  0.019\pm  0.003^{+  0.001}_{-  0.003}$ \\
   &  &  & 165 & 250 & 210 & 6.05 & 12.20 & 11 & $  0.014\pm  0.004^{+  0.002}_{-  0.004}$ \\
   &  &  & 165 & 250 & 210 &12.20 & 33.00 & 22 & $  0.007\pm  0.001^{+  0.001}_{-  0.001}$ \\
 \hline
\end{tabular}
\end{center}
\caption{
        The values of $d\sigma_{\rm diff}^{\gamma^*p}/dM_X$
        measured with the LPS method with the bin ranges indicated.
        The first and second error values represent
        the statistical and systematic uncertainties, respectively.
The normalisation uncertainty related to the luminosity measurement is
not included in the systematic uncertainty.
}
\label{tbl:lps2}
\end{table}

\begin{table}
\begin{center}
\begin{tabular}{|c|c|c||c|c|c||c|c|c||c|}
\hline
$Q^{2}_{\rm min}$ & $Q^{2}_{\rm max}$ & $ Q^{2} $ &
$W_{\rm min}$ & $W_{\rm max}$ & $W $ &
$M_{X, \rm min}$ & $M_{X,\rm max}$ & $M_X$ &
{$\frac{d\sigma_{\rm diff}^{\gamma^*p}}{dM_X}$} \\
 \hline
\multicolumn{3}{ |c||}{(GeV$^{2}$)} &
\multicolumn{3}{ c||}{(GeV)} &
\multicolumn{3}{ c||}{(GeV)} &
($\mu$b/GeV) \\
 \hline
 \hline
  0.220 & 0.324 & 0.27 &  90 & 120 & 104 & 3.00 &  6.05 & 4.26 &$  0.490\pm  0.022^{+  0.065}_{-  0.036}$ \\
   &  &  & 120 & 150 & 134 &  &   &  &$  0.557\pm  0.025^{+  0.056}_{-  0.036}$ \\
   &  &  & 150 & 180 & 164 &  &   &  &$  0.612\pm  0.029^{+  0.068}_{-  0.027}$ \\
   &  &  & 180 & 200 & 190 &  &  &  &$  0.698\pm  0.040^{+  0.057}_{-  0.028}$ \\
   &  &  & 200 & 220 & 210 &  &  &  &$  0.768\pm  0.047^{+  0.080}_{-  0.053}$ \\
 \hline
 \hline
  0.220 & 0.324 & 0.27 &  90 & 120 & 104 & 6.05 & 12.20 & 8.58 &$  0.200\pm  0.010^{+  0.039}_{-  0.019}$ \\
   &  &  & 120 & 150 & 134 &  &  &  &$  0.218\pm  0.010^{+  0.033}_{-  0.017}$ \\
   &  &  & 150 & 180 & 164 &  &  &  &$  0.246\pm  0.012^{+  0.024}_{-  0.010}$ \\
   &  &  & 180 & 200 & 190 &  &  &  &$  0.259\pm  0.015^{+  0.015}_{-  0.013}$ \\
   &  &  & 200 & 220 & 210 &  &  &  &$  0.291\pm  0.019^{+  0.025}_{-  0.022}$ \\
 \hline
 \hline

  0.324 & 0.476 & 0.39 &  90 & 120 & 104 & 3.00 &  6.05 & 4.26 &$  0.433\pm  0.019^{+  0.034}_{-  0.019}$ \\
   &  &  & 120 & 150 & 134 &  &   &  &$  0.455\pm  0.021^{+  0.039}_{-  0.032}$ \\
   &  &  & 150 & 180 & 164 &  &   &  &$  0.531\pm  0.027^{+  0.050}_{-  0.034}$ \\
   &  &  & 180 & 200 & 190 &  &   &  &$  0.599\pm  0.037^{+  0.047}_{-  0.038}$ \\
 \hline
 \hline
  0.324 & 0.476 & 0.39 &  90 & 120 & 104 & 6.05 & 12.20 & 8.58 &$  0.171\pm  0.008^{+  0.020}_{-  0.008}$ \\
   &  &  & 120 & 150 & 134 &  &  &  &$  0.186\pm  0.009^{+  0.019}_{-  0.012}$ \\
   &  &  & 150 & 180 & 164 &  &  &  &$  0.200\pm  0.010^{+  0.016}_{-  0.010}$ \\
   &  &  & 180 & 200 & 190 &  &  &  &$  0.238\pm  0.015^{+  0.013}_{-  0.011}$ \\
 \hline
 \hline
  0.476 & 0.700 & 0.58 &  90 & 120 & 104 & 3.00 &  6.05 & 4.26 &$  0.373\pm  0.019^{+  0.031}_{-  0.018}$ \\
   &  &  & 120 & 150 & 134 &  &   &  &$  0.411\pm  0.022^{+  0.038}_{-  0.033}$ \\
   &  &  & 150 & 180 & 164 &  &   &  &$  0.432\pm  0.026^{+  0.048}_{-  0.036}$ \\
 \hline
 \hline
  0.476 & 0.700 & 0.58 &  90 & 120 & 104 & 6.05 & 12.20 & 8.58 &$  0.149\pm  0.008^{+  0.017}_{-  0.011}$ \\
   &  &  & 120 & 150 & 134 &  &  &  &$  0.166\pm  0.009^{+  0.014}_{-  0.012}$ \\
   &  &  & 150 & 180 & 164 &  &  &  &$  0.162\pm  0.010^{+  0.010}_{-  0.010}$ \\
 \hline
\end{tabular}
\end{center}
\caption{
        The diffractive cross-sections $d\sigma_{\rm diff}^{\gamma^*p}/dM_{X}$
        measured with the $M_X^2$ method with the bin ranges indicated.
        The first and second error values represent
        the statistical and systematic uncertainties, respectively.
        The $\pm 21$\% systematic uncertainty due to the double-dissociation
        correction is not included in the systematic uncertainty, 
nor is the normalisation 
uncertainty related to the luminosity measurement.
}
\label{tab:xsection}
\end{table}

\begin{thebibliography}{99}
\bibitem{pdbc} P.D.B. Collins, {\it An Introduction to Regge Theory and
High Energy 
Physics}, Cambridge University Press, Cambridge (1977).

\bibitem{dl} 
A. Donnachie and P.V. Landshoff, Nucl. Phys. {\bf B244}, 322 (1984);\\
A. Donnachie and P.V. Landshoff, Phys. Lett. {\bf B296}, 227 (1992);\\
see also: \\
J.R. Cudell, K. Kang and S.K. Kim, Phys. Lett. {\bf B395}, 311 (1997); \\
Particle Data Group, D.E. Groom {\it et al.}, Eur. Phys. J. {\bf C15}, 1
(2000).

\bibitem{h1mx} H1 Collab., C. Adloff {\it et al.},
              Z. Phys. {\bf C74}, 221 (1997).
\bibitem{zeusphp} ZEUS Collab., J. Breitweg {\it et al.},
                Z. Phys. {\bf C75}, 421 (1997).

                
\bibitem{review_halina} See, for example, H. Abramowicz,
 Int. J. Mod. Phys. {\bf A15S1}, 495 (2000) and references therein.

\bibitem{h1diff} H1 Collab., C. Adloff {\it et al.},
                Z. Phys. {\bf C76}, 613 (1997).
\bibitem{zeusdiff94} ZEUS Collab., J. Breitweg {\it et al.},
                    Eur. Phys. J. {\bf C6}, 43 (1999).

\bibitem{flux} L.N. Hand, Phys. Rev. {\bf 129}, 1834 (1963).

\bibitem{ingelman} G. Ingelman and K. Janson-Prytz, {\it Proceedings of
the Workshop ``Physics at HERA''}, Volume 1, W. Buchm\"uller, G. Ingelman
(eds.), DESY (1992), p.~233;\\
G. Ingelman and K. Prytz, Z. Phys. {\bf C58}, 285 (1993).

\bibitem{rsmall}
M.F. McDermott and G. Briskin, {\it Proceedings of the Workshop
                 ``Future Physics at HERA''}, Volume 2, G.Ingelman, 
                 A. De Roeck and R. Klanner (eds.), DESY (1996), p.~691,
                 and references therein.

\bibitem{bluebook} ZEUS Collab., U.~Holm (ed.), {\it The ZEUS Detector,} 
       Status Report, (unpublished), DESY (1993);\\
  {\tt http://www-zeus.desy.de/bluebook/bluebook.html}~.

\bibitem{ctd} N. Harnew {\it et al.}, Nucl. Instr. and Meth. {\bf A279}, 290 (1989);\\
     B. Foster {\it et al.}, Nucl. Phys. Proc. Suppl. {\bf B32}, 181 (1993);\\
     B. Foster {\it et al.}, Nucl. Inst. and Meth. {\bf A338}, 254 (1994).

\bibitem{cal} M. Derrick {\it et al.}, Nucl. Instr. and Meth. {\bf A309},77 (1991);\\
   A. Andresen {\it et al.}, Nucl. Instr. and Meth. {\bf A309}, 101 (1991);\\
   A. Caldwell {\it et al.}, Nucl. Instr. and Meth. {\bf A321}, 356 (1992);\\
   A. Bernstein {\it et al.}, Nucl. Instr. and Meth. {\bf A336}, 23 (1993).

 \bibitem{zeusbpc}
   ZEUS Collab., J. Breitweg {\it et al.}, Phys. Lett. {\bf B407}, 432
(1997).

\bibitem{zeusbpt}
        ZEUS Collab., J. Breitweg {\it et al.}, Phys. Lett. {\bf B487},
53 (2000).

\bibitem{srtd} A. Bamberger {\it et al.}, Nucl. Instr. and Meth. {\bf A382},
419 (1996).
        
\bibitem{lpsrho} ZEUS Collab., M. Derrick {\it et al.}, Z. Phys. 
{\bf C73}, 253 (1997).

\bibitem{DA}
S. Bentvelsen, J. Engelen and P. Kooijman, in {\it Proceedings of the
Workshop on Physics at HERA, Oct. 1991}, Volume 1, W. Buchm\"uller and G.
Ingelman (eds.),  DESY (1992), p. 23;\\
K.C. H\"oger, ibid., p. 43.

\bibitem{gb} G. Briskin, PhD Thesis, Tel Aviv University (1998),
             DESY-Thesis-1998-036.
             
\bibitem{jb} F. Jacquet and A. Blondel, {\it Proceedings of the  
Workshop ``Study for an $ep$ Facility for Europe''}, U. Amaldi (ed.),
DESY 79-048 (1979), p.~391.

\bibitem{mk} M. Kasprzak, PhD Thesis, Warsaw University (1996),
             DESY F35D-96-16.

\bibitem{masahide} M. Inuzuka, PhD Thesis, University of Tokyo (1999), 
KEK Report 99-9.

\bibitem{heracles} K. Kwiatkowski, H. Spiesberger and H.-J. M\"ohring,
Comput. Phys. Commun. {\bf 69}, 155 (1992). 
          
\bibitem{rapgap} H. Jung, DESY Report 93-182 (1993).

\bibitem{ingsch}
G. Ingelman and P. Schlein, Phys. Lett. {\bf B152}, 256 (1985).
                                                                                
\bibitem{geant}
GEANT 3.13, R. Brun et al., CERN DD/EE/84-1 (1987).

\bibitem{sinkus} H. Abramowicz, A. Caldwell and R. Sinkus, Nucl. Instr. 
and Meth. {\bf A365}, 508 (1995).

\bibitem{lps_f2d3}
ZEUS Collab., J. Breitweg {\it et al.}, Eur. Phys. J. {\bf  C1}, 81
(1998).

\bibitem{svd} A. Hoecker and V. Kartvelishvili, Nucl. Instr. and Meth.
{\bf A372}, 469 (1996).

\bibitem{vf} 
 See e.g.:\\ 
U. Amaldi, M. Jacob and G. Matthiae, Ann. Rev. Nucl. Sci. {\bf 26}, 385
(1976);\\
G. Cohen-Tannoudji, D. Levy and M. Souza, Nucl. Phys. {\bf B129}, 286
(1977);\\ 
G. Alberi and G. Goggi, Phys. Rep. {\bf74}, 1 (1981);\\
K. Goulianos, Phys. Rep. {\bf101}, 169 (1983);\\
M. Kamran, Phys. Rep. {\bf108}, 275 (1984);\\
N.P. Zotov and V.A. Tsarev, Sov. Phys. Uspekhi {\bf 31}, 119 (1988);\\
G. Giacomelli, Int. J. Mod. Phys. {\bf A5}, 223 (1990).

\bibitem{zeuspheno} ZEUS Collab., J.Breitweg {\it et al.},
                  Eur. Phys. J. {\bf C7}, 609 (1999).

\bibitem{tslope_php} ZEUS Collab., J. Breitweg {\it et al.},
                 Eur. Phys. J. {\bf C2}, 237 (1998).

\bibitem{allm97} H.Abramowicz and A.Levy, DESY Report 97-251 (1997).

\bibitem{F29697} ZEUS Collab., S. Chekanov {\it et al.},
                Eur. Phys. J. {\bf C21}, 443 (2001).

\bibitem{h1php} H1 Collab., C. Adloff {\it et al.},
                Z.Phys. {\bf C74}, 221 (1997).


\bibitem{reviews} 
M. W\"usthoff and A. D. Martin, J. Phys. {\bf G25}, R309 (1999);\\
A. Hebecker,  Phys. Rep. {\bf 331}, 1 (2000);
              Acta Phys. Polon. {\bf B30} (1999) 3777;\\
K. Golec--Biernat and  M. W\"usthoff, Eur. Phys. J. {\bf C20}, 313
(2001).



\bibitem{bekw}  J. Bartels {\it et al.},
                Eur. Phys. J. {\bf C7}, 443 (1999).


\end{thebibliography}
\end{document}